\documentclass[10pt,letterpaper]{article}

\usepackage[margin=1in]{geometry}
\usepackage{amsmath,amssymb,amsfonts,mathtools,bm}
\usepackage{microtype}
\usepackage{enumitem}
\usepackage[compress]{cite}\usepackage{xcolor}
\usepackage{hyperref}

\hypersetup{
  colorlinks=true,
  linkcolor=blue,
  citecolor=blue,
  urlcolor=cyan,
  pdftitle={On the Spontaneous symmetry breaking of Conformal Gravity to Weyl and Einstein Gravities},
  pdfauthor={Partha Nandi, Gerasimos Kouniatalis, Stelios Stefas, George Zoupanos}
}

\allowdisplaybreaks

\newcommand{\Tr}{\operatorname{tr}}
\newcommand{\dd}{\mathrm{d}}

\newcommand{\cE}{\mathcal{E}}
\newcommand{\YM}{\mathrm{YM}}
\newcommand{\CG}{\mathrm{CG}}
\newcommand{\WG}{\mathrm{WG}}
\newcommand{\EG}{\mathrm{EG}}
\newcommand{\PM}{\mathrm{PM}}
\newcommand{\SO}{\mathrm{SO}}
\newcommand{\SU}{\mathrm{SU}}
\newcommand{\ii}{\mathrm{i}}

\begin{document}

\begin{center}
{\LARGE On the Spontaneous symmetry breaking of Conformal Gravity to\\[0.25em]
Weyl and Einstein Gravities\par}
\vspace{1.4em}
{\large Partha Nandi$^{1,2}$, Gerasimos Kouniatalis$^{3,4}$, Stelios Stefas$^{4}$ and George Zoupanos$^{4,5,6,7}$\par}
\vspace{1.6em}
{\itshape
$^{1}$ National Institute for Theoretical and Computational Sciences (NITheCS), South Africa\\
$^{2}$ Department of Physics, Stellenbosch University, Stellenbosch 7600, South Africa\\
$^{3}$National Observatory of Athens, Lofos Nymfon, 11852 Athens, Greece\\
$^{4}$Physics Department, National Technical University of Athens, 157 80, Zografou, Greece\\
$^{5}$Max-Planck Institut f\"ur Physik, Boltzmannstr. 8, 85 748 Garching/Munich, Germany\\
$^{6}$Universit\"at Hamburg, Luruper Chaussee 149, 22 761 Hamburg, Germany\\
$^{7}$Deutsches Elektronen-Synchrotron DESY, Notkestrasse 85, 22 607, Hamburg, Germany\par
\vspace{1.2em}
E-mails: \href{mailto:partha.nandi@nithecs.ac.za}{partha.nandi@nithecs.ac.za},
\href{mailto:gkouniatalis@noa.gr}{gkouniatalis@noa.gr},
\href{mailto:dstefas@mail.ntua.gr}{dstefas@mail.ntua.gr},\\
\href{mailto:george.zoupanos@cern.ch}{george.zoupanos@cern.ch}
}
\end{center}

\begin{abstract}

Does the special-conformal connection remain an auxiliary field when a spontaneously broken conformal gauge theory is promoted to a fully dynamical Yang--Mills--Higgs theory, and under what conditions does its elimination yield Weyl gravity? We address these questions within an \(\mathrm{SO}(2,4)\) gauge-theoretic formulation of gravity. Revisiting the gravitational sector selected by spontaneous symmetry breaking through an adjoint \(15\)-plet, we retain the special-conformal connection as an independent field and demonstrate that, on a source-free projected branch, its algebraic equation of motion determines it in terms of the Schouten tensor. Eliminating this connection yields the four-dimensional Weyl-squared action, up to the Euler density, without imposing the Schouten relation as an additional geometrical constraint. We then extend the analysis to the standard local Yang--Mills--Higgs theory and show that the special-conformal connection remains auxiliary, owing to the structure of the invariant bilinear form of \(\mathfrak{so}(2,4)\), which prevents the appearance of an independent kinetic operator. The additional gauge and scalar sectors can modify its algebraic solution, so the exact Weyl reduction is restricted to the source-free branch. We further investigate the linearised spectrum of the reduced Weyl theory around de Sitter spacetime, identifying massless and partially massless spin--2 branches, with the latter occurring at \(m_{\mathrm{PM}}^2=2\Lambda/3\). Finally, we discuss the subsequent symmetry breaking induced by an antisymmetric scalar \(6\)-plet and its role in connecting the Weyl-gravity phase to Einstein gravity.
\end{abstract}

\section{Introduction}

The unification of all fundamental interactions has been a longstanding
objective in theoretical physics, motivating diverse approaches and
geometrical frameworks over the past century. The pioneering proposal
of Kaluza and Klein~\cite{ref1,ref2} initiated the exploration of
additional spacetime dimensions as a possible means of unifying
fundamental interactions. This idea was subsequently revived and
generalized during the 1960s and 1970s~\cite{ref3,ref5,ref4},
developed more systematically within Coset Space Dimensional Reduction
(CSDR)~\cite{ref6,ref7,ref8,ref9,ref10,ref11,ref12,ref13,ref14,ref15,ref16},
and extensively pursued within the framework of superstring
theories~\cite{ref17,ref19,ref18}.

A complementary direction towards unification emerges from the
gauge-theoretic formulation of fundamental interactions. The fact
that the Standard Model of elementary-particle physics, Grand Unified
Theories, and gravity can all be formulated in terms of gauge
symmetries and connections naturally raises the question of whether
these interactions can be accommodated within a common gauge
structure~\cite{ref20,ref21,ref23,ref22,ref24,ref25,ref28,ref26,ref27,ref29}.
The gauge approach to gravity predates superstring theory and received
renewed attention during the development of supergravity~\cite{ref30,ref31}.
Subsequent developments have extended this perspective to noncommutative
gravity and generalized geometrical frameworks
~\cite{ref32,ref33,ref35,ref34,ref36,ref37,ref38,ref39}.

Beyond the broader question of unification, gauge-theoretic
formulations of gravity raise a more specific issue: how does the
gravitational sector emerge from an enlarged spacetime symmetry, and
what dynamical role is played by the gauge fields associated with
the broken generators? The gauge formulation of gravity has developed
through the works of Weyl~\cite{ref40,ref41}, Utiyama~\cite{ref20},
Kibble~\cite{ref21} and Sciama~\cite{ref22}. Weyl introduced local
scale invariance as a guiding principle for relating gravitation
and electromagnetism, while Utiyama showed that gravity could be
formulated as a gauge theory of the Lorentz group $\SO(1,3)$,
although the vierbein had to be introduced separately. Kibble and
Sciama subsequently developed this approach by gauging the full
Poincar\'e group. Further investigations of Poincar\'e gauge gravity
and Riemann--Cartan geometry have explored its geometrical and
canonical structures, particularly the role of torsion, the
relationship between spacetime symmetries and Hamiltonian gauge
transformations, and possible cosmological applications
~\cite{Banerjee:2009zz,PhysRevD.98.083506}. Stelle and West
~\cite{ref28,ref29} subsequently introduced constructions based on
the de Sitter $\SO(1,4)$ and anti-de Sitter $\SO(2,3)$ groups,
employing spontaneous symmetry breaking to recover Lorentz invariance.
The gauging of the conformal group $\SO(2,4)$ provided a further
framework for investigating the geometrical and dynamical structure
of gravity, including conformal and Weyl gravities~\cite{ref42,ref43},
their supersymmetric extensions~\cite{ref30,ref42}, and more recent
fuzzy-gravity constructions~\cite{ref33,ref35,ref34,ref36,ref37,ref38,ref39}.
Related modern formulations of Weyl-conformal geometry and quadratic
Weyl gravity further emphasize the gauge-theoretic role of local
scale symmetry and its geometrical realizations~\cite{ref75,ref76}.
In a different context, the extended $\SO(2,4)$ algebra has also
been employed to construct Lorentz-covariant quantum mechanics and
deformed phase-space structures with applications to noncommutative
geometry~\cite{Nandi:2023tfq}.

An important distinction among gauge-theoretic approaches to gravity
concerns the geometrical structure assumed at the outset and the
role played by spontaneous symmetry breaking in establishing
spacetime geometry. In the Stelle--West construction~\cite{ref28,ref29},
one starts with a de Sitter or anti-de Sitter gauge connection
together with an internal vector subject to a fixed-norm constraint.
The vector and its gauge-covariant derivative allow one to construct
a gauge-covariant metric-like geometrical object without independently
postulating a spacetime metric at the outset. The constraint selects
a symmetry-breaking orbit, while in a particular symmetry-breaking
gauge the covariant derivative of the internal vector reduces to
the vierbein, recovering the familiar metric geometry. Dynamical
extensions of this framework further promote the constrained vector
to a Higgs field with its own kinetic term and potential.

A different perspective is adopted in recent
pregeometric construction~\cite{Addazi:2024rzo,Capozziello:2026pys}.
There, the underlying manifold is differentiable
but is not endowed with an independent
pseudo-Riemannian metric. Instead, the emergence
of metric geometry is intended to follow dynamically
from spontaneous symmetry breaking driven by the
Wilczek term \cite{Wilczek:1998}. The present work adopts a different
geometrical starting point. We formulate the theory
on a four-dimensional pseudo-Riemannian spacetime
manifold, with the vierbein already identified as
a component of the conformal gauge connection,
from which the spacetime metric is constructed.
A dynamical adjoint-valued Higgs field in the
$15$ representation is then introduced to implement
spontaneous symmetry breaking and select the
gravitational sector. Thus, unlike our earlier
pregeometric approach, the objective here is not
the emergence of metric geometry itself, but
the dynamical determination of gravitational
sectors within an already geometrical conformal
gauge theory.

A more direct and ambitious possibility is to place
gravity and the internal gauge symmetries of particle
physics within a single enlarged tangent group,
as explored in early works~\cite{Weinberg:1984,ref44,ref45}
and subsequently revisited in
Refs.~\cite{ref38,ref43,ref46,ref47,ref48,ref49,ref50,ref51,ref52,ref53,ref54,ref55}.
The underlying observation is that the tangent group
of a curved spacetime need not have the same dimension
as the spacetime manifold. A tangent group of dimension
larger than four can therefore be gauged over an ordinary
four-dimensional manifold, allowing internal interactions
to be incorporated alongside gravity. Several techniques
developed for genuinely higher-dimensional CSDR
constructions~\cite{ref6,ref7,ref8,ref10,ref11,ref12,ref13,ref14,ref15,ref16,ref56}
carry over to this framework, as do the associated
restrictions on realistic chiral spectra
~\cite{ref11,ref56}. Along these lines, unification
of the conformal gauge group with internal interactions
was achieved in Refs.~\cite{ref38,ref43,ref51,ref53,ref54,ref55,ref66}
and subsequently extended to the noncommutative
case~\cite{ref39}.

The symmetry-breaking structure of unified gauge theories can
also influence the effective cosmological background. In
particular, while the gravitational contribution may favor
an anti-de Sitter (AdS) phase, vacuum contributions associated
with the breaking of the internal gauge symmetry can modify
the effective cosmological constant and potentially favor
a de Sitter (dS) background. This provides a broader
motivation for considering the dS case explicitly in our
subsequent analysis of the propagating spin--2 spectrum.

Returning to conformal gravity (CG), obtained by gauging
$\SO(2,4)$, Refs.~\cite{ref43,ref53,ref55,ref58} developed
a symmetry-breaking construction based on an auxiliary
scalar in the adjoint $15$ representation. Its vacuum
alignment along the dilatation generator selects the
gravitational sector through a MacDowell--Mansouri-type
projection. In the present construction, we extend this
framework to a dynamical setting by promoting the gauge
and scalar sectors to a dynamical theory, with the
vacuum expectation value of the adjoint-valued Higgs
field chosen as $\Phi_0=-2D/m$, where $D$ denotes the
dilatation generator.

In the earlier non-dynamical construction, different
gravitational theories were obtained by imposing
geometrical relations between the vierbein and the
gauge field associated with the special-conformal
generators. In particular, a proportionality relation
leads to an Einstein--Hilbert contribution with a
cosmological term, whereas the curvature-dependent
Schouten relation yields Weyl gravity. These results
provide the starting point for the present investigation
and raise a more specific question: can the geometrical
relation underlying the Weyl-gravity branch emerge
dynamically from the field equations of the fully
dynamical conformal gauge theory, rather than being
imposed as an independent prescription?

Our first step is to revisit precisely this
Stage-A gravitational sector while retaining
the broken special-conformal connection
$b_\mu{}^a$ off shell, rather than fixing it
from the outset. On the source-free projected
branch defined below, its own algebraic field
equation determines its relation to the
spacetime curvature and yields the Schouten
relation. Substituting this solution into the
projected action reduces it exactly to
four-dimensional Weyl gravity, up to the
topological Euler contribution. Thus, the
geometrical relation that previously had to
be selected can, on this branch, be obtained
from the variational principle. This is the
first central distinction from the earlier
non-dynamical construction: the Schouten
relation is derived dynamically rather than
imposed as an independent geometrical
prescription. The qualification is important.
The exact Weyl reduction holds on the specified
source-free projected branch, where the complete
additional contributions to the special-conformal
connection equation from the Yang--Mills and
scalar kinetic sectors vanish; it is not
asserted for arbitrary configurations of
the full dynamical theory.

Our second step is to establish whether the
auxiliary character of $b_\mu{}^a$ persists
once the gauge and scalar sectors are made
dynamical. We show that this is indeed the
case in the standard local $\SO(2,4)$
Yang--Mills--Higgs theory. The adjoint-invariant
Yang--Mills bilinear contains a mixed
translation--special-conformal curvature block,
but no independent special-conformal square.
Consequently, the derivative of $b_\mu{}^a$
occurs only linearly and can be removed by
integration by parts, leaving an algebraic
field equation. The additional dynamical
sectors may nevertheless contribute algebraic
sources to this equation, without generating
an independent kinetic term for the
special-conformal connection. Thus, the exact
Schouten solution characterizes the
source-free projected gravitational branch,
whereas the auxiliary nature of $b_\mu{}^a$
is a more general structural property of
the standard local theory. This distinction
is important for understanding the role of
the special-conformal connection: it helps
determine the gravitational configuration
without introducing an additional independent
propagating gauge field.

Having obtained the projected Weyl-gravity
action, we investigate its propagating
spin--2 spectrum on a de Sitter background.
The fourth-order metric equation separates
into massless and partially massless (PM)
spin--2 branches. The PM branch occurs at
$m_{\rm PM}^{2}=2\Lambda/3$, precisely the
Higuchi value~\cite{ref57,ref67,ref68,ref69}.
For comparison, a conventional massive
Fierz--Pauli spin--2 field has a healthy
five-helicity representation for
$m_{\rm FP}^{2}>2\Lambda/3$, whereas at equality
its helicity-zero mode is removed by an additional
scalar gauge symmetry, yielding the partially
massless representation.

In our construction, the PM value characterizes
the corresponding branch of the projected
Weyl-gravity spectrum. Following symmetry
breaking, the physical spin--2 excitation is
genuinely massive, with its mass determined by
the symmetry-breaking scale. Its mass can coincide
with the Higuchi value, corresponding to the
partially massless limit, or exceed it, in which
case it describes a healthy massive Fierz--Pauli
field with five helicities. The spin--2 excitation
belongs to the propagating metric sector and should
not be confused with a massive excitation of the
auxiliary special-conformal connection. This
clarifies the relation between the Weyl-gravity
spectrum, spontaneous symmetry breaking and the
Higuchi bound.

Finally, we investigate a subsequent
symmetry-breaking stage driven by a scalar
in the antisymmetric $6$ representation.
While the first breaking selects the
gravitational sector through the adjoint
$15$-plet, the second vacuum breaks the
remaining dilatation symmetry and lifts
the degeneracy between the translation
and special-conformal gauge directions.
It separates the corresponding gravitational
fields into distinct light and heavy
combinations. At energies well below the
heavy mass, the heavy combination can be
integrated out, while the light combination
provides the effective vierbein of the
low-energy gravitational phase. This
provides a dynamical symmetry-breaking
route towards Einstein gravity, distinct
from imposing a proportionality relation
between the vierbein and the special-conformal
connection as an independent geometrical
prescription. The two breaking stages
therefore offer a framework for understanding
how different gravitational phases can
arise from an underlying conformal gauge
theory.

The structure of the paper is as follows. Section~2 recalls the non-dynamical $15$-plet construction and fixes the conventions. Section~3 presents the full dynamical theory in both the physical and complex bases. Section~4 establishes the auxiliary nature of the special-conformal connection in the standard local Yang--Mills--Higgs sector. Section~5 derives the Weyl-gravity action on the source-free projected branch. Section~6 discusses the de Sitter spectrum, the Higuchi point, and the partially massless sector. Section~7 investigates the second symmetry-breaking stage and the continuation towards Einstein gravity. We conclude in Sec.~8.

\section{Background: conformal gauge gravity and symmetry breaking with a non-dynamical 15-plet}
In this section we summarize the gauge-theoretic construction of gravity based on $\SO(2,4)$ and the way Weyl and Einstein gravities (WG and EG, respectively) were obtained from it in Refs.~\cite{ref43,ref53,ref55,ref58}, where symmetry breaking was driven by a non-dynamical auxiliary scalar in the adjoint $15$ representation. We then state the ingredients that are changed in the present paper.

\subsection{The conformal gauge theory}
The fifteen generators of $\mathfrak{so}(2,4)$ split into Lorentz generators $M_{ab}$, translations $P_a$, special conformal generators $K_a$, and the dilatation $D$, with non-vanishing commutators
\begin{align}
[M_{ab},M_{cd}]&=\eta_{ac}M_{bd}-\eta_{ad}M_{bc}-\eta_{bc}M_{ad}+\eta_{bd}M_{ac},\nonumber\\
[M_{ab},P_c]&=\eta_{ac}P_b-\eta_{bc}P_a,\qquad
[M_{ab},K_c]=\eta_{ac}K_b-\eta_{bc}K_a,\nonumber\\
[D,P_a]&=P_a,\qquad [D,K_a]=-K_a,\nonumber\\
[P_a,K_b]&=2(\eta_{ab}D-M_{ab}).
\label{eq:alg}
\end{align}
Gauging this algebra introduces the connection
\begin{equation}
A_\mu=\frac12\omega_\mu{}^{ab}M_{ab}+e_\mu{}^aP_a+b_\mu{}^aK_a+\chi_\mu D,
\label{eq:connection}
\end{equation}
where $e_\mu{}^a$ is the vierbein, $\omega_\mu{}^{ab}$ the spin connection, $b_\mu{}^a$ the gauge field of special conformal transformations, and $\chi_\mu$ the dilatation gauge field. We use $\chi_\mu$ uniformly throughout the paper. The field strength
\begin{equation}
F_{\mu\nu}=\frac12R_{\mu\nu}{}^{ab}M_{ab}+\widetilde R_{\mu\nu}{}^aP_a+R_{\mu\nu}{}^aK_a+R_{\mu\nu}^{(D)}D
\end{equation}
has components
\begin{align}
R_{\mu\nu}{}^{ab}&=R_{\mu\nu}^{(0)ab}-8e_{[\mu}{}^{[a}b_{\nu]}{}^{b]},\label{eq:RabComp}\\
\widetilde R_{\mu\nu}{}^a&=T_{\mu\nu}^{(0)a}(e)+2\chi_{[\mu}e_{\nu]}{}^a,\label{eq:RPComp}\\
R_{\mu\nu}{}^a&=T_{\mu\nu}^{(0)a}(b)-2\chi_{[\mu}b_{\nu]}{}^a,\label{eq:RKComp}\\
R_{\mu\nu}^{(D)}&=\partial_\mu\chi_\nu-\partial_\nu\chi_\mu+4e_{[\mu}{}^ab_{\nu]a},
\label{eq:RDComp}
\end{align}
where
\begin{align}
R_{\mu\nu}^{(0)ab}&=\partial_\mu\omega_\nu{}^{ab}-\partial_\nu\omega_\mu{}^{ab}-\omega_\mu{}^{ac}\omega_{\nu c}{}^b+\omega_\nu{}^{ac}\omega_{\mu c}{}^b,\\
T_{\mu\nu}^{(0)a}(e)&=\partial_\mu e_\nu{}^a-\partial_\nu e_\mu{}^a-\omega_\mu{}^{ab}e_{\nu b}+\omega_\nu{}^{ab}e_{\mu b},
\end{align}
and $T^{(0)}(b)$ is the same expression with $e\rightarrow b$.

\subsection{Breaking with a non-dynamical 15}
A convenient way to implement spontaneous symmetry breaking in the conformal gauge theory is to introduce an auxiliary scalar field $\phi$ transforming in the adjoint representation of $\mathrm{SU}(4)$, whose dimension is $15$. We refer to this field as the adjoint $15$-plet. It is non-dynamical in the sense that no independent kinetic term is included for it; rather, it serves as an auxiliary field whose fixed norm and vacuum orientation select the symmetry-breaking direction. Here, the use of $\mathrm{SU}(4)\simeq\mathrm{SO}(6)$ is understood at the level of the Lie algebras and through their common complexification with the conformal algebra $\mathfrak{so}(2,4)$. A parity-preserving action quadratic in the curvature is
constructed by coupling the auxiliary scalar $\phi$ to the
curvature two-form. To fix the norm of $\phi$, we introduce
a mass parameter $m$ and a Lagrange multiplier $\lambda$,
which enforces the corresponding constraint. The resulting
action is
\begin{equation}
S=a_{\CG}\int \dd^4x\,\Tr\!\left[
\epsilon^{\mu\nu\rho\sigma}m\phi F_{\mu\nu}F_{\rho\sigma}
+\lambda\left(\phi^2-m^{-2}{\bf 1}_4\right)
\right].
\label{eq:aux15action}
\end{equation}
Here, $m$ sets the fixed norm of the auxiliary field,
while $a_{\CG}$ is the dimensionless conformal-gravity
coupling that determines the overall normalization of
the action. The field $\lambda$ acts as a Lagrange
multiplier enforcing the constraint
$\phi^2=m^{-2}{\bf 1}_4$.

We now choose the gauge in which the vacuum expectation
value of $\phi$ is aligned with the dilatation generator,
\begin{equation}
\phi=\phi_0=-2m^{-1}D.
\end{equation}
With this choice, the curvature-dependent part of the
action becomes
\begin{equation}
S=-2a_{\CG}\int \dd^4x\,
\Tr\!\left(\epsilon^{\mu\nu\rho\sigma}
F_{\mu\nu}F_{\rho\sigma}D\right).
\end{equation}
This vacuum alignment implements the spontaneous
symmetry breaking. In particular, the adjoint
representation branches under
\begin{equation}
\mathrm{SU}(4)\longrightarrow
\mathrm{SU}(2)\times\mathrm{SU}(2)\times U(1)
\end{equation}
as
\begin{equation}
15=(3,1)_0+(1,3)_0+(1,1)_0
+(2,2)_{+2}+(2,2)_{-2}.
\end{equation}
The two triplets correspond to the Lorentz generators,
the singlet to the dilatation generator $D$, and the
two $(2,2)$ representations to the translation and
special-conformal generators, $P_a$ and $K_a$,
respectively. Consequently, the $P_a$ and $K_a$
directions are broken, whereas the Lorentz and
dilatation generators, $M_{ab}$ and $D$, span the
stabilizer algebra of the chosen vacuum.

Expanding the field strengths and evaluating the relevant traces, the surviving gravitational part is
\begin{equation}
S=\frac{a_{\CG}}{4}\int\dd^4x\,\epsilon^{\mu\nu\rho\sigma}\epsilon_{abcd}
R_{\mu\nu}{}^{ab}R_{\rho\sigma}{}^{cd}.
\label{eq:projected}
\end{equation}
In the non-dynamical construction the curvatures $\widetilde R_{\mu\nu}{}^a$, $R_{\mu\nu}{}^a$, and $R_{\mu\nu}^{(D)}$ do not appear in Eq.~\eqref{eq:projected}. They can therefore be consistently constrained to vanish, together with $\chi_\mu$. In particular, $R_{\mu\nu}^{(D)}=0$ gives
\begin{equation}
e_\mu{}^ab_{\nu a}-e_\nu{}^ab_{\mu a}=0,
\end{equation}
so that the spacetime tensor $b_{\mu\nu}\equiv e_{\nu a}b_\mu{}^a$ is symmetric on that constrained branch.

\subsection{Einstein and Weyl gravity}
In the non-dynamical treatment $b_\mu{}^a$ was not varied; instead one selected a relation between $b_\mu{}^a$ and the vierbein compatible with the preceding constraint. Two choices were studied.

\paragraph{Case 1: $b_\mu{}^a=a\,e_\mu{}^a$.}
Substitution into Eq.~\eqref{eq:projected} gives a Gauss--Bonnet term, an Einstein--Hilbert term, and a cosmological-constant term~\cite{ref59}. For an appropriate sign of $a$ this reproduces General Relativity in an AdS phase. This branch is recalled here only as part of the earlier non-dynamical construction; in the dynamical treatment below it is not used as an explicit Weyl-to-Einstein breaking prescription.

\paragraph{Case 2: the Schouten relation.}
A curvature-dependent relation between $b_\mu{}^a$ and $e_\mu{}^a$ gives the Weyl action~\cite{ref30,ref60}. In the conventions used below, it can be written as
\begin{equation}
b_{\mu\nu}=\frac12P_{\mu\nu},\qquad
P_{\mu\nu}\equiv\frac12\left(R_{\mu\nu}-\frac16Rg_{\mu\nu}\right),
\label{eq:historicalSchouten}
\end{equation}
where an opposite sign for $b_\mu{}^a$ is frequently used in the conformal-gauge literature. The corresponding curvature becomes the Weyl tensor and Eq.~\eqref{eq:projected} reduces, up to the Euler density, to
\begin{equation}
S_W=2a_{\CG}\int\dd^4x\sqrt{-g}\left(R_{\mu\nu}R^{\mu\nu}-\frac13R^2\right),
\label{eq:oldWeyl}
\end{equation}
which is the scale-invariant four-dimensional Weyl action.

\subsection{What changes in the present work}

The preceding construction treats the scalar as a Lagrange-multiplier
field: it has no kinetic term, its vacuum value is enforced by a
constraint, and the only gauge-field dynamics relevant to
Eq.~\eqref{eq:projected} arise from the projected $\epsilon$-type
term. Here the theory is instead taken to be local and dynamical,
\begin{equation}
S=S_{\rm MM}+S_{\YM}+S_{\Phi},
\label{eq:fullActionIntro}
\end{equation}
with
\begin{align}
S_{\rm MM}
&=\widehat a_{\CG}m\int
\Tr\!\left(\Phi_{15}\,F\wedge F\right),
\label{eq:SMM}\\
S_{\YM}
&=-\frac{1}{2g_{\YM}^{2}}
\int\kappa(F\wedge\star F),
\label{eq:SYMdef}\\
S_{\Phi}
&=-\frac12\int\dd^4x\sqrt{-g}\,
\Tr\!\left(D_\mu\Phi_{15}D^\mu\Phi_{15}\right)
-\int\dd^4x\sqrt{-g}\,V(\Phi_{15}),
\label{eq:SPhidef}
\end{align}
where $\kappa$ is the symmetric adjoint-invariant bilinear form
and $V$ is a gauge-invariant potential whose broken branch contains
the adjoint vacuum. The metric entering $S_{\YM}$ and $S_\Phi$ is
constructed from the vierbein, see Appendix~\ref{app:metric}.
We denote the coefficient of the unprojected MacDowell--Mansouri
term by $\widehat a_{\CG}$; after insertion of the adjoint vacuum
and evaluation of the fixed generator trace, the resulting overall
factor is absorbed into the effective Stage-A coefficient
$a_{\CG}$ used in the projected action below. The essential changes
are now precise: the scalar is dynamical; the Yang--Mills term
contains all conformal curvatures; the curvatures omitted from
Eq.~\eqref{eq:projected} can no longer be discarded merely because
they were absent from the projected term; and the status of
$b_\mu{}^a$ has to be determined from the complete local action
rather than by an imposed relation.

\section{Dynamical \texorpdfstring{$\SO(2,4)$}{SO(2,4)}
theory in physical and complex variables}
\label{sec:complex}

\paragraph{Physical Lorentzian fields and curvatures.}

The physical Lorentzian gauge algebra is the real conformal algebra
\begin{equation}
\mathfrak{so}(2,4)\simeq\mathfrak{su}(2,2),
\end{equation}
and the physical gauge fields are $\omega^{ab}$, $e^a$, $b^a$,
and $\chi$. In differential-form notation,
\begin{equation}
A=\frac12\omega^{ab}M_{ab}
+\chi D+e^aP_a+b^aK_a,
\end{equation}
and
\begin{equation}
F=\frac12R^{ab}M_{ab}
+F_DD+T^aP_a+S^aK_a.
\end{equation}
The curvatures are
\begin{align}
T^a
&=D^{(\omega)}e^a+\chi\wedge e^a,
\label{eq:Tform}\\
S^a
&=D^{(\omega)}b^a-\chi\wedge b^a,
\label{eq:Sform}\\
F_D
&=\dd\chi+2e^a\wedge b_a,
\label{eq:FDform}\\
R^{ab}
&=R^{ab}(\omega)-4e^{[a}\wedge b^{b]}.
\label{eq:Rform}
\end{align}
Only $S^a$ contains a derivative of the special-conformal
connection. In components,
\begin{equation}
S_{\mu\nu}{}^a
=2D^{(\omega)}_{[\mu}b_{\nu]}{}^a
-2\chi_{[\mu}b_{\nu]}{}^a.
\label{eq:Scomp}
\end{equation}
Thus an independent kinetic operator for $b_\mu{}^a$ could only
originate from an $S^2=F^{(K)2}$ block.

\paragraph{Invariant bilinear in the physical basis.}

Let $\kappa(X,Y)$ be a symmetric adjoint-invariant bilinear form,
\begin{equation}
\kappa([X,Y],Z)+\kappa(Y,[X,Z])=0.
\label{eq:invbil}
\end{equation}
Taking $X=D$ and using Eq.~\eqref{eq:alg},
\begin{align}
0
&=\kappa([D,P_a],P_b)
+\kappa(P_a,[D,P_b])
=2\kappa(P_a,P_b),\\
0
&=\kappa([D,K_a],K_b)
+\kappa(K_a,[D,K_b])
=-2\kappa(K_a,K_b),
\end{align}
so
\begin{equation}
\kappa(P_a,P_b)=0,\qquad
\kappa(K_a,K_b)=0.
\label{eq:nullblocks}
\end{equation}
Lorentz covariance permits the mixed block
\begin{equation}
\kappa(P_a,K_b)=c\,\eta_{ab},
\qquad c\neq0.
\label{eq:PKpair}
\end{equation}
For the simple algebra $\mathfrak{so}(2,4)$, the symmetric
adjoint-invariant bilinear form is unique up to an overall
multiplicative normalization. We fix this normalization by
choosing $c=1$, so that
\begin{equation}
\kappa(P_a,K_b)=\eta_{ab}.
\end{equation}
The remaining components of the bilinear form then follow
from adjoint invariance:
\begin{align}
\kappa(D,D)&=\frac12,\\
\kappa(M_{ab},D)&=0,\\
\kappa(M_{ab},M_{cd})
&=-\frac12
\left(\eta_{ac}\eta_{bd}-\eta_{ad}\eta_{bc}\right).
\label{eq:remainingbilinear}
\end{align}
Therefore, with
\begin{equation}
F_{P,K}=T^aP_a+S^aK_a,
\end{equation}
one obtains
\begin{equation}
\kappa(F_{P,K},F_{P,K})
=2T^a\!\cdot S_a.
\label{eq:TSphysical}
\end{equation}
There is no $T^2$ block and, crucially, no $S^2$ block.
Physically, this means that the Yang--Mills invariant does not
provide the special-conformal connection with its own wave
operator: the field can enter the local dynamics, but it cannot
propagate an independent $K_a$ gauge excitation through a
standard $(\partial b)^2$ term. This absence follows from the
structure of the invariant bilinear form and is not a
tunable choice.

\paragraph{Complex \texorpdfstring{$P$--$K$}{P--K} coordinates.}

The same result can be displayed in complex notation without
introducing any new physical field. Define
\begin{equation}
Q_a\equiv\frac{P_a-\ii K_a}{\sqrt2},
\qquad
Q_a^{\star}\equiv\frac{P_a+\ii K_a}{\sqrt2},
\label{eq:Qdef}
\end{equation}
and the corresponding complex one-forms
\begin{equation}
Z^a\equiv\frac{e^a+\ii b^a}{\sqrt2},
\qquad
Z^{\star a}\equiv\frac{e^a-\ii b^a}{\sqrt2}.
\label{eq:Zdef}
\end{equation}
Here $\star$ denotes Lorentzian reality conjugation, not the
Hodge dual. The inverse relations are
\begin{equation}
e^a=\frac{Z^a+Z^{\star a}}{\sqrt2},
\qquad
b^a=\frac{Z^a-Z^{\star a}}{\ii\sqrt2},
\end{equation}
and indeed
\begin{equation}
Z^aQ_a+Z^{\star a}Q_a^{\star}
=e^aP_a+b^aK_a.
\end{equation}
Introduce
\begin{equation}
W^a\equiv\frac{T^a+\ii S^a}{\sqrt2}
=D^{(\omega)}Z^a+\chi\wedge Z^{\star a}.
\label{eq:Wdef}
\end{equation}
The bilinear becomes
\begin{equation}
\kappa(Q_a,Q_b)=-\ii\eta_{ab},
\qquad
\kappa(Q_a^{\star},Q_b^{\star})
=+\ii\eta_{ab},
\qquad
\kappa(Q_a,Q_b^{\star})=0.
\label{eq:Qbil}
\end{equation}
The invariant contraction is therefore complex bilinear,
not Hermitian. Writing
\begin{equation}
F_{Q\oplus Q^{\star}}
\equiv W^aQ_a+W^{\star a}Q_a^{\star},
\end{equation}
we have
\begin{align}
W\!\cdot W
&=\frac12\left(T^2-S^2+2\ii T\!\cdot S\right),\\
W^{\star}\!\cdot W^{\star}
&=\frac12\left(T^2-S^2-2\ii T\!\cdot S\right),
\end{align}
and therefore
\begin{align}
\kappa(F_{Q\oplus Q^{\star}},F_{Q\oplus Q^{\star}})
&=-\ii W\!\cdot W
+\ii W^{\star}\!\cdot W^{\star}
\nonumber\\
&=2T\!\cdot S.
\label{eq:complexcancel}
\end{align}
There is no cross bilinear in this expression: the two displayed
terms come from the diagonal $Q$--$Q$ and
$Q^{\star}$--$Q^{\star}$ blocks, while the cross block vanishes
identically, $\kappa(Q_a,Q_b^{\star})=0$. The $T^2-S^2$ pieces
cancel exactly. By contrast,
\begin{equation}
W\!\cdot W^{\star}=\frac12(T^2+S^2)
\end{equation}
does contain $S^2$, but this Hermitian contraction would multiply
the vanishing cross block and is therefore not selected by the
ordinary adjoint-invariant Yang--Mills bilinear.

\section{The special-conformal connection is auxiliary in the
standard local theory}
\label{sec:auxproof}

\paragraph{Yang--Mills derivative Hessian.}

With the normalization of the invariant bilinear form fixed
in Sec.~\ref{sec:complex}, the Yang--Mills action is
\begin{equation}
S_{\YM}
=-\frac{1}{4g_{\YM}^{2}}
\int\dd^4x\sqrt{-g}\,
\kappa(F_{\mu\nu},F^{\mu\nu}).
\label{eq:YMaction}
\end{equation}
Using
\begin{align}
\kappa(M_{ab},M_{cd})
&=-\frac12
(\eta_{ac}\eta_{bd}-\eta_{ad}\eta_{bc}),\\
\kappa(D,D)&=\frac12,\\
\kappa(P_a,K_b)&=\eta_{ab},
\end{align}
the contraction of the full curvature is
\begin{align}
\kappa(F_{\mu\nu},F^{\mu\nu})
={}&-\frac14
R_{\mu\nu}{}^{ab}R^{\mu\nu}{}_{ab}
+\frac12 F_{D\mu\nu}F_D{}^{\mu\nu}
\nonumber\\
&+2T_{\mu\nu}{}^aS^{\mu\nu}{}_a.
\label{eq:fullYMcontraction}
\end{align}
Substituting this result into Eq.~\eqref{eq:YMaction}, we obtain
\begin{equation}
S_{\YM}
=\int\dd^4x\sqrt{-g}\left[
\alpha R_{\mu\nu}{}^{ab}R^{\mu\nu}{}_{ab}
+\beta F_{D\mu\nu}F_D{}^{\mu\nu}
+\lambda T_{\mu\nu}{}^aS^{\mu\nu}{}_a
\right],
\label{eq:YMsectors}
\end{equation}
where
\begin{equation}
\alpha=\frac{1}{16g_{\YM}^{2}},
\qquad
\beta=-\frac{1}{8g_{\YM}^{2}},
\qquad
\lambda=-\frac{1}{2g_{\YM}^{2}}.
\label{eq:YMcoefficients}
\end{equation}
Thus, the relative coefficients are fixed by the
$\mathfrak{so}(2,4)$ algebra and the chosen normalization of
its invariant bilinear form. In particular,
\begin{equation}
\alpha:\beta:\lambda=1:-2:-8,
\end{equation}
so they are not independent Yang--Mills couplings.

The Lorentz curvature~\eqref{eq:Rform} and $F_D$ in
Eq.~\eqref{eq:FDform} contain $b^a$ only algebraically.
The only derivative-containing term is the mixed $T\cdot S$
piece. Using Eq.~\eqref{eq:Scomp} and the antisymmetry of
$T^{\mu\nu}$,
\begin{align}
\mathcal L_{TS}
&=\sqrt{-g}\,\lambda T_a{}^{\mu\nu}
\left(
2D^{(\omega)}_{[\mu}b_{\nu]}{}^a
-2\chi_{[\mu}b_{\nu]}{}^a
\right)\nonumber\\
&=2\sqrt{-g}\,\lambda T_a{}^{\mu\nu}
D^{(\omega)}_\mu b_\nu{}^a
-2\sqrt{-g}\,\lambda T_a{}^{\mu\nu}
\chi_\mu b_\nu{}^a.
\end{align}
Thus the complete dependence on $\partial_\mu b_\nu{}^a$
is affine,
\begin{equation}
\mathcal L_{\partial b}
=2\sqrt{-g}\,\lambda T_a{}^{\mu\nu}
\partial_\mu b_\nu{}^a,
\end{equation}
and the derivative Hessian vanishes identically,
\begin{equation}
\frac{\partial^2\mathcal L}
{\partial(\partial_\mu b_\nu{}^a)\,
\partial(\partial_\rho b_\sigma{}^b)}=0.
\label{eq:derHessian}
\end{equation}
Equivalently, the velocity Hessian is
\begin{equation}
H^{ij}_{~ab}\equiv
\frac{\partial^2\mathcal L}
{\partial\dot b_i{}^a\partial\dot b_j{}^b}=0,
\end{equation}
and
\begin{equation}
\pi^{i}_{~a}
\equiv\frac{\partial\mathcal L}{\partial\dot b_i{}^a}
=2\sqrt{-g}\,\lambda T_a{}^{0i}
\end{equation}
contains no invertible dependence on $\dot b_i{}^a$. Thus
$b_\mu{}^a$ has no independent Yang--Mills kinetic matrix.
The physical content of this degeneracy is stronger than a
mere choice of variables: specifying initial data for the
remaining fields does not require an additional freely
propagating $b$-sector velocity.

\paragraph{Removing the unique derivative by integration by parts.}

A vanishing Hessian establishes degeneracy, but the stronger
auxiliary-field statement follows because the remaining first
derivative can be removed. Varying only $b$ while holding
$e$, $\omega$, and $\chi$ fixed,
\begin{equation}
\delta_b S_{\mu\nu}{}^a
=2D^{(\omega)}_{[\mu}\delta b_{\nu]}{}^a
-2\chi_{[\mu}\delta b_{\nu]}{}^a.
\end{equation}
For
\begin{equation}
S_{TS}
=\lambda\int\dd^4x\sqrt{-g}\,
T_a{}^{\mu\nu}S_{\mu\nu}{}^a,
\end{equation}
integration by parts gives
\begin{equation}
\delta_bS_{TS}
=-2\lambda\int\dd^4x\sqrt{-g}
\left(
D^{(\omega)}_\mu T_a{}^{\mu\nu}
+\chi_\mu T_a{}^{\mu\nu}
\right)\delta b_\nu{}^a,
\label{eq:ibpvar}
\end{equation}
up to a boundary term. No derivative of $b$ remains in its
Euler--Lagrange equation. At the action level,
\begin{equation}
\begin{aligned}
\int\sqrt{-g}\,
2\lambda T_a{}^{\mu\nu}
D^{(\omega)}_\mu b_\nu{}^a
={}&-\int\sqrt{-g}\,
2\lambda b_\nu{}^a
D^{(\omega)}_\mu T_a{}^{\mu\nu}\\
&+\text{boundary}.
\end{aligned}
\end{equation}
After this step every occurrence of $b_\mu{}^a$ in the
standard local Yang--Mills action is algebraic.

\paragraph{Higgs and projected gravitational sectors.}

For an adjoint $15$-plet,
\begin{equation}
D_\mu\Phi_{15}
=\partial_\mu\Phi_{15}-\ii g[A_\mu,\Phi_{15}],
\end{equation}
while for an antisymmetric $6=\wedge^24$ written in complex
$\SU(2,2)$ notation,
\begin{equation}
D_\mu\Sigma_6
=\partial_\mu\Sigma_6
-\ii g\left(A_\mu\Sigma_6+\Sigma_6A_\mu^{T}\right).
\label{eq:6covder}
\end{equation}

In either case, the gauge connection enters the covariant
derivative algebraically. In particular, the special-conformal
component $b_\mu{}^aK_a$ enters $D_\mu\Phi_{15}$ and
$D_\mu\Sigma_6$ without derivatives of $b_\mu{}^a$.
Consequently, the scalar kinetic terms can generate both terms
quadratic in $b_\mu{}^a$ and terms linear in $b_\mu{}^a$.
They therefore contribute, in general, to both the algebraic
coefficient and the linear source in the $b_\mu{}^a$ equation.
The scalar potential $V(\Phi_{15})$, being independent of the
gauge connection, gives no direct contribution to this equation.

The same algebraic dependence occurs in the projected
MacDowell--Mansouri sector, since
\begin{equation}
R^{ab}=R^{ab}(\omega)-4e^{[a}\wedge b^{b]},
\qquad
\delta_bR^{ab}=-4e^{[a}\wedge\delta b^{b]},
\end{equation}
contains no derivative of $\delta b$.

The remaining Yang--Mills curvature blocks also depend
algebraically on $b_\mu{}^a$, apart from the derivative in
$S^a$, which has already been removed from the action by
integration by parts. Thus, after this integration by parts,
every occurrence of $b_\mu{}^a$ in the standard local action
is algebraic.

Consequently, after the mixed $T\cdot S$ term has been
integrated by parts, the complete standard local action has
the schematic but exact algebraic structure
\begin{equation}
S=S_0[e,\omega,\chi,\Phi_r]
+\int\dd^4x\sqrt{-g}
\left(
\frac12b_I\mathcal M^{IJ}b_J+b_I\mathcal J^I
\right),
\label{eq:generalAlg}
\end{equation}
where $I=(\mu,a)$ is a compound index.

Here $S_0$ is independent of $b_\mu{}^a$, while
$\mathcal M^{IJ}$ and $\mathcal J^I$ contain contributions
from all sectors of the action. In particular, we may
decompose them as

\begin{align}
\mathcal{M}^{IJ}
&=
\mathcal{M}_{\mathrm{proj}}^{IJ}
+\mathcal{M}_{\mathrm{YM}}^{IJ}
+\mathcal{M}_{\Phi}^{IJ},
\label{eq:Mdecomposition}
\\
\mathcal{J}^{I}
&=
\mathcal{J}_{\mathrm{proj}}^{I}
+\mathcal{J}_{\mathrm{YM}}^{I}
+\mathcal{J}_{\Phi}^{I}.
\label{eq:Jdecomposition}
\end{align}

The scalar contributions $\mathcal M_\Phi^{IJ}$ and
$\mathcal J_\Phi^I$ arise from the scalar kinetic terms and
need not vanish when the scalar is evaluated at its vacuum
expectation value. The potential contributes no direct
$b$-dependence. Neither $\mathcal M^{IJ}$ nor $\mathcal J^I$
contains derivatives of $b_\mu{}^a$; however, $\mathcal J^I$
may contain derivatives of the other fields, such as
$D_\mu^{(\omega)}T_a{}^{\mu\nu}$.

Hence the complete equation obtained by varying with respect
to $b_I$ is
\begin{equation}
\mathcal M^{IJ}b_J+\mathcal J^I=0.
\label{eq:generalbeom}
\end{equation}

The decomposition above makes an important distinction
explicit: the absence of a differential operator establishes
that $b_\mu{}^a$ is auxiliary, but it does not determine its
algebraic solution. In particular, the scalar kinetic
contribution can modify both the coefficient multiplying
$b_\mu{}^a$ and the corresponding linear source. Thus, even
when the torsion vanishes, the scalar sector can shift the
algebraic equation for the special-conformal connection.

Relative to the projected gravitational term it is useful
to write
\begin{equation}
\mathcal M=\mathcal M_{\rm proj}+\Delta\mathcal M,
\qquad
\mathcal J=\mathcal J_{\rm proj}+\Delta\mathcal J,
\end{equation}
where
\begin{equation}
\Delta\mathcal M
=\mathcal M_{\YM}+\mathcal M_\Phi,
\qquad
\Delta\mathcal J
=\mathcal J_{\YM}+\mathcal J_\Phi.
\end{equation}

The additional terms therefore include the complete
algebraic contributions from both the Yang--Mills and scalar
kinetic sectors. In particular, setting only the linear
scalar source $\mathcal J_\Phi^I$ to zero would not be
sufficient to remove the scalar contribution from the
$b$ equation, since the quadratic term
$\mathcal M_\Phi^{IJ}b_J$ may remain nonzero.

If $\mathcal M$ is invertible on the branch of interest,
\begin{equation}
b_I=-(\mathcal M^{-1})_{IJ}\mathcal J^J,
\end{equation}
which is a local algebraic elimination. If $\mathcal M$ is
singular, the corresponding components impose constraints
rather than acquiring an independent kinetic equation.
In neither case does the standard local sector generate
a propagating $b$ mode by itself.

The auxiliary statement fails only if the theory is enlarged
beyond this standard contraction. In the complex notation
the missing operator is isolated by
\begin{equation}
W\!\cdot W^{\star}
-\mathrm{Re}(W\!\cdot W)
=S^2=F^{(K)2}.
\end{equation}
Adding an explicit term proportional to $S^2$, or a
scalar-dependent gauge-kinetic tensor
$G_{AB}(\Phi)F^AF^B$ that generates a non-zero effective
$K$--$K$ block, produces a genuine $(\partial b)^2$
contribution and invalidates the auxiliary reduction.
Such terms are additional effective-field-theory assumptions
and are not generated by the ordinary constant
adjoint-invariant bilinear.

\section{Exact Stage-A reduction to Weyl gravity}
\label{sec:weylreduction}

We now specialize to the projected gravitational sector
selected by the adjoint breaking. First let us make the
projection from the MacDowell--Mansouri term explicit.
In the broken phase choose the gauge
\[
\langle\Phi_{15}\rangle=v_{15}D,
\qquad
F\big|_{\rm L}=\frac12F^{ab}M_{ab}.
\]
With the generator normalization fixed in Sec.~2, the
Lorentz trace obeys
\[
\Tr\!\left(DM_{ab}M_{cd}\right)
=-\frac12\epsilon_{abcd},
\]
the same trace identity underlying
Eqs.~\eqref{eq:aux15action}--\eqref{eq:projected}.
Consequently the Lorentz part of $S_{\rm MM}$ is precisely
an $\epsilon_{abcd}F^{ab}F^{cd}$ term. We define the
effective Stage-A coefficient $a_{\CG}$ by absorbing into
it the bare coefficient $\widehat a_{\CG}$, the factor
$m v_{15}$ and the fixed numerical factors associated with
the chosen two-form convention. Equivalently, the
normalization is fixed by
\[
\left.S_{\rm MM}\right|_{\langle\Phi_{15}\rangle,\,\rm L}
\equiv
\frac{a_{\CG}}{4}\int\dd^4x\,
\epsilon^{\mu\nu\rho\sigma}\epsilon_{abcd}
F_{\mu\nu}{}^{ab}F_{\rho\sigma}{}^{cd}.
\]
For the historical normalization
$\phi_0=-2m^{-1}D$ this definition reproduces
Eq.~\eqref{eq:projected}. This projection concerns the
MacDowell--Mansouri term only; it does not remove the
Yang--Mills or scalar sectors from the full action.

The auxiliary theorem of Sec.~\ref{sec:auxproof} is therefore
logically more general than the Schouten solution derived
below. In particular, the mixed Yang--Mills term contributes
\begin{equation}
\mathcal J_{a}{}^{\nu}\big|_{TS}
=-2\lambda
\left(
D^{(\omega)}_\mu T_a{}^{\mu\nu}
+\chi_\mu T_a{}^{\mu\nu}
\right)
=
\frac{1}{g_{\YM}^{2}}
\left(
D^{(\omega)}_\mu T_a{}^{\mu\nu}
+\chi_\mu T_a{}^{\mu\nu}
\right).
\label{eq:bsourceTS}
\end{equation}

For the explicit Stage-A reduction, we now impose a further
condition on the additional sectors. The projected
MacDowell--Mansouri action is retained as the gravitational
action whose $b_\mu{}^a$ equation determines the Schouten
relation. The Yang--Mills and scalar kinetic sectors, however,
can modify this equation through additional algebraic terms.
We therefore restrict the following reduction to configurations
on which their combined contribution to the $b_\nu{}^a$
Euler--Lagrange equation vanishes. We define this source-free
projected branch by

\begin{equation}
\left.\mathcal E_a{}^\nu\right|_{\rm extra}
\equiv
\left.
\frac{1}{\sqrt{-g}}
\frac{\delta\!\left(S_{\YM}+S_{\Phi}\right)}
{\delta b_\nu{}^a}
\right|_{\rm branch}
=0.
\label{eq:sourcefreebranch}
\end{equation}

This is a condition on the complete additional
Euler--Lagrange contribution, not merely on its part linear
in $b_\mu{}^a$. In the decomposition of
Eqs.~\eqref{eq:Mdecomposition}--\eqref{eq:Jdecomposition},
it requires

\begin{equation}
\left.
\left(
\Delta\mathcal M^{IJ}b_J+\Delta\mathcal J^I
\right)
\right|_{\rm branch}=0.
\label{eq:completeExtraCondition}
\end{equation}

Here the projected MacDowell--Mansouri contribution is
excluded from the additional terms. In particular, the
scalar kinetic contribution is included in
Eq.~\eqref{eq:sourcefreebranch}; it is not assumed to vanish
simply because the scalar is evaluated at its vacuum
expectation value.

The mixed Yang--Mills term gives the specific
derivative-generated contribution displayed in
Eq.~\eqref{eq:bsourceTS}. Thus, the torsionless condition
$T^a=0$ eliminates this particular contribution. It does
not, by itself, establish Eq.~\eqref{eq:sourcefreebranch}:
the algebraic contributions from the remaining Yang--Mills
curvature blocks and, importantly, from the scalar kinetic
term must also satisfy the branch condition. Whether this
happens for the chosen scalar vacuum and couplings requires
evaluation of the complete additional $b_\nu{}^a$ equation.

If Eq.~\eqref{eq:sourcefreebranch} is not satisfied,
$b_\mu{}^a$ remains auxiliary by Sec.~\ref{sec:auxproof},
but its algebraic solution is shifted and need not equal
the Schouten tensor.

The equations below therefore establish an exact Weyl
reduction of the projected gravitational sector and
coincide with the full $b$ equation only on configurations
satisfying Eq.~\eqref{eq:sourcefreebranch}. They are not
claimed to provide a universal Schouten solution for
arbitrary configurations of the full dynamical theory.

\paragraph{Off-shell form of the projected action.}

To obtain the projected action without imposing any algebraic
condition on the special-conformal connection, we introduce
the spacetime tensor
\begin{equation}
b_{\mu\nu}\equiv e_{\nu a}b_\mu{}^a,
\qquad
b\equiv g^{\mu\nu}b_{\mu\nu},
\end{equation}
and leave $b_{\mu\nu}$ completely unrestricted, in particular
without assuming symmetry under $\mu\leftrightarrow\nu$.
It is also convenient to define
\begin{equation}
X_{\mu\nu}{}^{ab}
\equiv e_{[\mu}{}^{[a}b_{\nu]}{}^{b]},
\end{equation}
in terms of which the projected curvature takes the form
\begin{equation}
F_{\mu\nu}{}^{ab}
=R_{\mu\nu}{}^{ab}-8X_{\mu\nu}{}^{ab}.
\end{equation}
The corresponding projected action is
\begin{equation}
S_A=\frac{a_{\CG}}{4}\int\dd^4x\,
\epsilon^{\mu\nu\rho\sigma}\epsilon_{abcd}
F_{\mu\nu}{}^{ab}F_{\rho\sigma}{}^{cd}.
\end{equation}

Throughout the calculation, we use the mostly-plus metric
signature and choose the orientation
\begin{equation}
\epsilon^{0123}=+1,
\qquad
\epsilon_{0123}=-1.
\end{equation}
With these conventions, the relevant contractions of the
Levi-Civita tensors are
\begin{align}
\frac14\epsilon^{\mu\nu\rho\sigma}\epsilon_{abcd}
R_{\mu\nu}{}^{ab}R_{\rho\sigma}{}^{cd}
&=-\sqrt{-g}\,\cE_4,
\label{eq:epsRR}\\
\epsilon^{\mu\nu\rho\sigma}\epsilon_{abcd}
R_{\mu\nu}{}^{ab}X_{\rho\sigma}{}^{cd}
&=4\sqrt{-g}\,G^{\mu\nu}b_{\mu\nu},
\label{eq:epsRX}\\
\epsilon^{\mu\nu\rho\sigma}\epsilon_{abcd}
X_{\mu\nu}{}^{ab}X_{\rho\sigma}{}^{cd}
&=2\sqrt{-g}
\left(b_{\mu\nu}b^{\nu\mu}-b^2\right).
\label{eq:epsXX}
\end{align}
Here,
\begin{equation}
\cE_4
=R_{\mu\nu\rho\sigma}R^{\mu\nu\rho\sigma}
-4R_{\mu\nu}R^{\mu\nu}+R^2
\end{equation}
denotes the four-dimensional Euler density, and
$G_{\mu\nu}=R_{\mu\nu}-\frac12Rg_{\mu\nu}$ is the
Einstein tensor.

Substituting these identities into the projected action,
we obtain its complete off-shell expression:
\begin{align}
S_A[g,b]
=S_{\rm GB}
-16a_{\CG}\int\dd^4x\sqrt{-g}\,
\Big[
G^{\mu\nu}b_{\mu\nu}
-2\left(b_{\mu\nu}b^{\nu\mu}-b^2\right)
\Big],
\label{eq:offShell}
\end{align}
where
\begin{equation}
S_{\rm GB}
=-a_{\CG}\int\dd^4x\sqrt{-g}\,\cE_4
\end{equation}
is the Gauss--Bonnet contribution.

An important feature of this expression is that the
quadratic dependence on $b_{\mu\nu}$ has the relative
trace structure familiar from the Fierz--Pauli mass term.
Here, however, this structure is generated directly by
the conformal curvature in the projected action; it is
not introduced through an independent massive-gravity
potential. Moreover, the expression has been obtained
without imposing symmetry on $b_{\mu\nu}$, allowing its
symmetric and antisymmetric components to be determined
independently through the subsequent field equations.

\paragraph{Algebraic field equation and Schouten tensor.}

Varying Eq.~\eqref{eq:offShell} with respect to a completely
general $b_{\mu\nu}$ gives
\begin{equation}
G^{\mu\nu}
-4\left(b^{\nu\mu}-bg^{\mu\nu}\right)=0.
\label{eq:bEOM}
\end{equation}
The antisymmetric part yields
\begin{equation}
b^{[\mu\nu]}=0,
\end{equation}
so symmetry follows dynamically rather than being imposed.
The symmetric equation is
\begin{equation}
G_{\mu\nu}-4(b_{\mu\nu}-bg_{\mu\nu})=0.
\end{equation}
Its trace gives
\begin{equation}
-R+12b=0,
\qquad
b=\frac{R}{12},
\end{equation}
and substitution back yields
\begin{equation}
b_{\mu\nu}
=\frac14\left(R_{\mu\nu}-\frac16Rg_{\mu\nu}\right)
=\frac12P_{\mu\nu}.
\label{eq:bSchouten}
\end{equation}

On the source-free projected branch specified in
Eqs.~\eqref{eq:bsourceTS}--\eqref{eq:sourcefreebranch},
this is an exact local algebraic solution of the full
$b$ equation because the complete additional
Euler--Lagrange contribution has been set to zero.
No differential equation for $b_{\mu\nu}$ has to be solved.
In physical terms, the broken special-conformal connection
acts as a curvature-adjusting auxiliary field: it follows
the local Ricci geometry through the Schouten tensor instead
of carrying an independent local wave. Away from this branch
the auxiliary character persists, but non-zero additional
algebraic contributions from the Yang--Mills and scalar
kinetic sectors shift the solution, and the pure Schouten
relation need not hold.

\paragraph{Exact metric reduction.}

Using Eq.~\eqref{eq:bSchouten},
\begin{align}
G^{\mu\nu}b_{\mu\nu}
&=\frac14\left(
R_{\mu\nu}R^{\mu\nu}-\frac13R^2
\right),\\
b_{\mu\nu}b^{\mu\nu}-b^2
&=\frac1{16}\left(
R_{\mu\nu}R^{\mu\nu}-\frac13R^2
\right).
\end{align}
Equation~\eqref{eq:offShell} then becomes
\begin{align}
S_A
&=-a_{\CG}\int\dd^4x\sqrt{-g}\left[
\cE_4
+2\left(R_{\mu\nu}R^{\mu\nu}-\frac13R^2\right)
\right]\nonumber\\
&=-a_{\CG}\int\dd^4x\sqrt{-g}\,
C_{\mu\nu\rho\sigma}C^{\mu\nu\rho\sigma}.
\label{eq:WeylExact}
\end{align}
Thus the Stage-A projected theory is exactly four-dimensional
Weyl gravity, with the Schouten relation obtained dynamically
from the equation of motion of the retained broken $K_a$
gauge field. The elimination therefore does not discard a
physical particle; it trades an auxiliary gauge connection
for the higher-derivative metric description in which the
same dynamics are encoded by the Weyl tensor.

\section{de Sitter expansion and partially massless factorization}
\label{sec:dS}
The exact nonlinear reduction is sufficient to establish the Weyl-gravity phase. To identify the propagating spin--2 branches we now expand that theory around a maximally symmetric de Sitter solution. The use of a maximally symmetric background is essential: it is the constant curvature that permits the fourth-order Bach operator to factorize into two second-order operators with a direct particle interpretation.

\paragraph{Background and auxiliary shift.}
We take
\begin{align}
\bar R_{\mu\nu}&=\Lambda\bar g_{\mu\nu},&
\bar R&=4\Lambda,\nonumber\\
\bar R_{\mu\nu\rho\sigma}
&=\frac{\Lambda}{3}
(\bar g_{\mu\rho}\bar g_{\nu\sigma}-\bar g_{\mu\sigma}\bar g_{\nu\rho}),&
\Lambda&>0.
\label{eq:dSbackgroundFull}
\end{align}
Equation~\eqref{eq:bSchouten} immediately fixes the background value of the auxiliary field,
\begin{equation}
\bar b_{\mu\nu}=\frac{\Lambda}{12}\bar g_{\mu\nu},
\qquad
\bar b=\frac{\Lambda}{3}.
\label{eq:bbackgroundFull}
\end{equation}
It is useful to isolate the fluctuation by defining
\begin{equation}
b_{\mu\nu}=\frac{\Lambda}{12}g_{\mu\nu}+\phi_{\mu\nu},
\label{eq:bshiftFull}
\end{equation}
and the cosmological Einstein tensor
\begin{equation}
\mathcal G_{\mu\nu}\equiv G_{\mu\nu}+\Lambda g_{\mu\nu}.
\label{eq:cosEinstein}
\end{equation}
Substitution into the exact auxiliary action gives, up to the Gauss--Bonnet term,
\begin{align}
S_A[g,\phi]
={}&\frac{4a_{\CG}\Lambda}{3}\int\dd^4x\sqrt{-g}\,(R-2\Lambda)
-16a_{\CG}\int\dd^4x\sqrt{-g}\,\mathcal G^{\mu\nu}\phi_{\mu\nu}\nonumber\\
&+32a_{\CG}\int\dd^4x\sqrt{-g}\,(\phi_{\mu\nu}\phi^{\mu\nu}-\phi^2).
\label{eq:shiftedExact}
\end{align}
No Einstein--Hilbert term has been added by hand in Eq.~\eqref{eq:shiftedExact}. It appears because the auxiliary field has been shifted around a nonzero de Sitter value; together with the $\phi_{\mu\nu}$ terms the complete expression remains exactly equivalent to the Weyl-squared action. Physically, this form separates the background curvature scale from the fluctuation that enforces the higher-derivative Weyl dynamics.

Writing $g_{\mu\nu}=\bar g_{\mu\nu}+h_{\mu\nu}$ and denoting by $\mathcal G^L_{\mu\nu}[h]$ the linearized cosmological Einstein tensor, the quadratic action is
\begin{align}
S_A^{(2)}={}&-\frac{2a_{\CG}\Lambda}{3}\int\dd^4x\sqrt{-\bar g}\,
h^{\mu\nu}\mathcal G^L_{\mu\nu}[h]
-16a_{\CG}\int\dd^4x\sqrt{-\bar g}\,\phi^{\mu\nu}\mathcal G^L_{\mu\nu}[h]\nonumber\\
&+32a_{\CG}\int\dd^4x\sqrt{-\bar g}\,(\phi_{\mu\nu}\phi^{\mu\nu}-\phi^2).
\label{eq:quadraticAux}
\end{align}
Variation with respect to $\phi_{\mu\nu}$ yields
\begin{equation}
\mathcal G^L_{\mu\nu}[h]-4(\phi_{\mu\nu}-\bar g_{\mu\nu}\phi)=0.
\label{eq:phiEom}
\end{equation}
Taking the trace and substituting back gives
\begin{equation}
\phi_{\mu\nu}=\frac14\left[\mathcal G^L_{\mu\nu}[h]-\frac13\bar g_{\mu\nu}\mathcal G^L[h]\right].
\label{eq:phiSolFull}
\end{equation}
Thus the fluctuation inherited from $b_{\mu\nu}$ remains auxiliary even on the curved background. Eliminating it produces
\begin{align}
S_{\rm eff}^{(2)}=-2a_{\CG}\int\dd^4x\sqrt{-\bar g}\left[
\mathcal G^L_{\mu\nu}\mathcal G_L^{\mu\nu}
-\frac13(\mathcal G^L)^2
+\frac{\Lambda}{3}h_{\mu\nu}\mathcal G_L^{\mu\nu}
\right].
\label{eq:quadraticEffFull}
\end{align}

\paragraph{Fourth-order equation and factorization.}
Variation of Eq.~\eqref{eq:quadraticEffFull} gives the linearized fourth-order metric equation. The spin--2 content is most transparent in the transverse-traceless sector,
\begin{equation}
\bar\nabla^\mu h^{\rm TT}_{\mu\nu}=0,
\qquad h^{\rm TT}=0.
\label{eq:TTconditions}
\end{equation}
On de Sitter,
\begin{equation}
\mathcal G^L_{\mu\nu}[h^{\rm TT}]
=-\frac12\left(\bar\square-\frac{2\Lambda}{3}\right)h^{\rm TT}_{\mu\nu}.
\label{eq:GTTfull}
\end{equation}
Substitution into the quadratic action, or equivalently variation of the Bach tensor, yields
\begin{equation}
\left(\bar\square-\frac{2\Lambda}{3}\right)
\left(\bar\square-\frac{4\Lambda}{3}\right)
h^{\rm TT}_{\mu\nu}=0.
\label{eq:BachFactorFull}
\end{equation}
The factorization is not an extra spectral assumption; it follows from the auxiliary Stage-A action after the algebraic field has been eliminated.

Define
\begin{equation}
L_0\equiv\bar\square-\frac{2\Lambda}{3},
\qquad
L_{\PM}\equiv\bar\square-\frac{4\Lambda}{3}.
\end{equation}
For $\Lambda\neq0$ (automatically satisfied on the de Sitter background with $\Lambda>0$ adopted here), any TT solution of $L_0L_{\PM}h^{\rm TT}=0$ may be decomposed into the two branches as
\begin{equation}
h^{(0)}_{\mu\nu}=-\frac{3}{2\Lambda}L_{\PM}h^{\rm TT}_{\mu\nu},
\qquad
h^{(\PM)}_{\mu\nu}=\frac{3}{2\Lambda}L_0h^{\rm TT}_{\mu\nu},
\label{eq:projectors}
\end{equation}
so that $L_0h^{(0)}=0$ and $L_{\PM}h^{(\PM)}=0$. The first equation is the massless graviton equation. For comparison, a Fierz--Pauli field of mass $m$ satisfies in TT gauge
\begin{equation}
\left(\bar\square-\frac{2\Lambda}{3}-m^2\right)\varphi^{\rm TT}_{\mu\nu}=0.
\label{eq:FPdS}
\end{equation}
The second factor therefore corresponds to
\begin{equation}
m_{\PM}^{2}=\frac{2\Lambda}{3}.
\label{eq:PMmassFull}
\end{equation}
At precisely this value the scalar gauge transformation
\begin{equation}
\delta\varphi_{\mu\nu}
=\left(\bar\nabla_\mu\bar\nabla_\nu+\frac{\Lambda}{3}\bar g_{\mu\nu}\right)\alpha
\label{eq:PMgaugeFull}
\end{equation}
removes the helicity-zero mode~\cite{ref68,ref67,ref69}. A generic massive spin--2 field has five polarizations in four dimensions; the partially massless representation has only four, with helicities $\pm2$ and $\pm1$.

The physical point is that the quantity $m_{\PM}$ is not a separate mass assigned to the auxiliary connection. It is fixed by the curvature of the de Sitter background after the theory has already reduced to Weyl gravity. The broken $K_a$ connection supplies the auxiliary representation that generates the Weyl action; the propagating PM mode is a branch of the resulting metric equation.

\paragraph{Higuchi bound and the partially massless point.}
For a conventional Fierz--Pauli spin--2 field on de Sitter, the five-helicity massive representation is unitary for $m^2>2\Lambda/3$, while the helicity-zero mode has negative norm for $0<m^2<2\Lambda/3$~\cite{ref68,ref67,ref57}. At the saturation value $m^2=2\Lambda/3$ an additional scalar gauge symmetry removes the helicity-zero mode and the linear field is partially massless. Equation~\eqref{eq:PMmassFull} places the second spin--2 branch of the reduced Weyl theory exactly at this special point. No tuning of the spontaneous-symmetry-breaking scale to the Higuchi value is required for this Weyl branch: $m_{\PM}^2$ is fixed by the de Sitter curvature after the auxiliary field has been eliminated. If a broader completion instead contains a genuinely massive Fierz--Pauli spin--2 particle whose mass is set by an SSB scale, the healthy five-helicity region requires $m^2>2\Lambda/3$; equality corresponds specifically to the PM representation.

The de Sitter factorization therefore contains two massless helicities and four partially massless helicities, in agreement with the standard six local degrees of freedom of four-dimensional Weyl gravity~\cite{ref65,ref71}.

\section{Interpretation of the broken conformal sector and further breaking}
\label{sec:interpretation}
The symmetry-breaking interpretation is most transparent when the algebraic reduction and the actual symmetry-breaking sequence are kept separate. On the source-free projected branch, the dynamical reduction map is
\begin{equation}
K_a\ \longrightarrow\ b_\mu{}^a\ \longrightarrow\ b_{\mu\nu}
\ \xrightarrow{\rm algebraic\ EOM}\ \frac12P_{\mu\nu}
\ \longrightarrow\ C^2,
\label{eq:physicalchain}
\end{equation}
while the de Sitter massless/PM decomposition is a subsequent spectral property of the resulting Weyl theory rather than another symmetry-breaking step.
The field $b_{\mu\nu}$ is therefore not itself the physical partially massless field. Its role is to provide an off-shell representation of curvature whose exact elimination produces the higher-derivative metric theory. Only after this reduction does the de Sitter background split the metric excitations into the massless and PM representations.

This distinction also explains the Fierz--Pauli trace structure that appeared in Eq.~\eqref{eq:offShell}. The combination $b_{\mu\nu}b^{\nu\mu}-b^2$ is generated by the $SO(2,4)$ curvature algebra, but it is an algebraic auxiliary quadratic form, not a freely chosen nonlinear potential for an independently propagating massive spin--2 field~\cite{ref62,ref63,ref64}.. The resemblance to a Fierz--Pauli mass term identifies the tensor structure, not the dynamical status of the field.

The Stage-A Weyl phase is not the end of the symmetry-breaking chain. The constructions of Refs.~\cite{ref43,ref53,ref55} also contain a scalar in the antisymmetric/vector $6$ representation. In compact notation,
\begin{equation}
SO(6)\supset SO(5),\qquad 6=1+5,
\end{equation}
followed by
\begin{equation}
SO(5)\supset SU(2)\times SU(2),\qquad 5=(1,1)+(2,2),
\end{equation}
organizes the second breaking. The Lorentzian theory uses the corresponding $SU(2,2)$ reality condition on the same complexified representation.

For the antisymmetric scalar $\Sigma_6$,
\begin{equation}
D_\mu\Sigma_6=\partial_\mu\Sigma_6-\ii g(A_\mu\Sigma_6+\Sigma_6A_\mu^T),
\label{eq:6covderFinal}
\end{equation}
so the gauge field $b_\mu{}^a$ appears without derivatives. Hence the second scalar sector does not invalidate the auxiliary theorem proved above.

There is a useful consistency check that should be distinguished
conceptually from the second spontaneous breaking. Consider the
Weyl-gravity phase on an Einstein background satisfying
\begin{equation}
\bar R_{\mu\nu}=\Lambda_{\rm bg}\,\bar g_{\mu\nu},
\qquad
\bar R=4\Lambda_{\rm bg}.
\end{equation}
The algebraic Schouten equation \eqref{eq:bSchouten} then gives
\begin{equation}
\bar b_{\mu\nu}
=\frac{\Lambda_{\rm bg}}{12}\,\bar g_{\mu\nu},
\qquad
\bar b_\mu{}^a=c\,\bar e_\mu{}^a,
\qquad
c=\frac{\Lambda_{\rm bg}}{12}.
\label{eq:EinsteinBackgroundb}
\end{equation}
Thus, on an Einstein background, the algebraic equation of motion
for the special-conformal connection automatically yields the
proportionality relation $b_\mu{}^a\propto e_\mu{}^a$ that also
characterizes the Einstein branch discussed in Sec.~2.3.
Conversely, imposing $b_\mu{}^a=c\,e_\mu{}^a$ directly in the
projected action produces an Einstein-Hilbert term together with
a cosmological term.

These two statements, however, have different logical starting
points. In the first, the proportionality between $b_\mu{}^a$
and $e_\mu{}^a$ follows from the algebraic equation of motion
evaluated on an assumed Einstein background; in the second, it is
imposed as an ansatz at the level of the action. Accordingly, the
parameter $c$ characterizing the action-level Einstein branch and
the background curvature $\Lambda_{\rm bg}$ are not identified
a priori. If both descriptions refer to the same Einstein
solution, their values are related by
$c=\Lambda_{\rm bg}/12$. The Einstein-background condition
therefore provides a consistency check on the algebraic
equation of motion, rather than an additional prescription for
spontaneous symmetry breaking from Weyl gravity to Einstein
gravity. The latter is associated with the separate 6-plet
breaking discussed above.
We therefore use the genuinely spontaneous route for the second stage: the Weyl-gravity phase is further broken to the Einstein-gravity phase by the scalar $6$-plet vacuum described above and in Refs.~\cite{ref43,ref53,ref55}. The broader relation between conformal and Einstein gravity on Einstein backgrounds has also been analyzed directly at the action level~\cite{ref77}. In this sense the Einstein-background relation \eqref{eq:EinsteinBackgroundb} is a consistency relation within the Weyl phase, whereas the $6$-plet vacuum is the symmetry-breaking mechanism that realizes $WG\to EG$. The construction sequence used throughout the paper is therefore
\begin{equation}
\SO(2,4)\xrightarrow{\langle15\rangle}
\text{Stage-A broken conformal phase}
\xrightarrow[\mathcal E_{b,\rm extra}=0]{\rm auxiliary\ elimination}\WG
\xrightarrow{\langle6\rangle}\EG.
\label{eq:fullbreakingchain}
\end{equation}

\section{Discussion and conclusions}
\label{sec:conclusions}

The present work develops an important step in the
gauge-theoretic construction of conformal gravity based on
$SO(2,4)$. Earlier constructions of Weyl or Einstein gravity
from conformal gauge theories employed curvature or gauge-field
constraints to select the gravitational sector
\cite{ref42,ref59,ref60,ref72}. Spontaneous symmetry breaking
provides an alternative framework for relating gauge-theoretic
gravity to its lower-symmetry phases
\cite{ref28,ref43,ref53,ref55,ref58}. However, an important question
remains: whether the relations between the conformal gauge
fields that lead to a particular gravitational theory should
be imposed as additional conditions or can instead be
derived from the field equations of the broken theory.
The present work addresses this question by retaining the
special-conformal connection as an independent field and
investigating its dynamical status in the local
Yang--Mills--Higgs formulation.

We have first revisited the gravitational sector selected
by the adjoint $15$-plet breaking considered in earlier
constructions \cite{ref43,ref53,ref55,ref58}. Rather than imposing
a relation between the vierbein and the special-conformal
connection, we retain the latter as an independent field
and derive its equation of motion. On the source-free
projected branch, the resulting algebraic equation fixes
the special-conformal connection to one-half of the
Schouten tensor \cite{ref61}. In particular, the symmetry
of the corresponding spacetime tensor follows from its
field equation rather than being assumed at the outset.
Substituting this solution back into the projected
gravitational action yields four-dimensional Weyl-squared
gravity, up to the Euler density. Thus, the Schouten
relation, previously used to select the Weyl-gravity
sector, is obtained here as a consequence of the
variational principle.

We have subsequently examined the status of the
special-conformal connection when the gauge and scalar
sectors are promoted to the standard local
Yang--Mills--Higgs theory. A central result of our
analysis is that the broken special-conformal connection
$b_\mu{}^a$ remains an auxiliary field. This follows
from the invariant bilinear structure of the
$\mathfrak{so}(2,4)$ gauge algebra: the Yang--Mills action
contains a mixed translation--special-conformal curvature
contribution but no independent special-conformal
curvature-square term. Consequently, derivatives of
$b_\mu{}^a$ occur only linearly and can be removed by
integration by parts, leaving an algebraic field equation.
The scalar kinetic sector likewise introduces no
independent kinetic operator for this connection.

It is important, however, to distinguish the general
auxiliary-field result from the particular Schouten
solution. The former holds within the standard local
Yang--Mills--Higgs theory considered here, whereas the
latter follows only when the complete additional
$b$-variation from the Yang--Mills and scalar sectors
vanishes. Non-vanishing additional algebraic
contributions can shift the solution for $b_\mu{}^a$,
without necessarily making it propagating. Thus, the
exact reduction to Weyl gravity is established on the
specified source-free projected branch, rather than
for arbitrary configurations of the complete theory.

Having established the projected Weyl-gravity phase,
we have examined its linearized spin--2 spectrum on a
de Sitter background. The corresponding metric equation
factorizes into massless and partially massless spin--2
branches, in agreement with the known structure of
four-dimensional Weyl gravity
\cite{ref67,ref69,ref57}. The partially massless branch
occurs at the Higuchi value
$m_{\rm PM}^{2}=2\Lambda/3$
\cite{ref68,ref67,ref69}. This excitation should not
be confused with a generic massive Fierz--Pauli
spin--2 field. A conventional massive spin--2 field
has a healthy five-helicity representation in de Sitter
space for $m^{2}>2\Lambda/3$, whereas equality
corresponds to the partially massless point, where
the helicity-zero mode is removed by an additional
scalar gauge symmetry \cite{ref68,ref67,ref69}.
The de Sitter curvature scale and the
symmetry-breaking mass scales therefore have
distinct origins and should not be identified
unless a complete vacuum solution relates them
dynamically. Moreover, the partially massless branch
belongs to the propagating metric spectrum of the
reduced Weyl theory; it does not represent an
independent propagating mode of the auxiliary
connection $b_\mu{}^a$.

In the wider unification programme in which the
conformal gauge sector and the $SO(10)$ internal
interactions are embedded in $SO(2,16)$
\cite{ref43,ref53,ref54,ref55,ref66,ref73}, de Sitter
backgrounds can arise naturally, making the
distinction between the curvature and
symmetry-breaking scales physically relevant.
We have also considered the subsequent breaking
induced by the antisymmetric scalar $6$-plet,
which provides a spontaneous continuation from
the Weyl phase towards Einstein gravity
\cite{ref43,ref53,ref55,ref58}. The associated
mass-matrix analysis distinguishes the two
combinations of translation and special-conformal
gauge directions and exhibits the mass splitting
induced by the second scalar vacuum. This
mechanism should be distinguished from the
proportional relation
$b_\mu{}^a\propto e_\mu{}^a$
obtained as a consistency condition on an
Einstein-type background within the Weyl phase:
the latter is not the symmetry-breaking
mechanism responsible for the subsequent
transition towards Einstein gravity.

The explicit representation-theoretic details
underlying the two symmetry-breaking stages
are collected in Appendix~A. In particular,
the compact generator basis and the corresponding
gauge-boson mass matrices are worked out there,
making the mass degeneracy after the first
breaking and its lifting after the second
breaking explicit. The compact representation
provides a convenient organization of the
mass calculation, while the Lorentzian
$SU(2,2)$ reality condition remains relevant
for the physical conformal theory.

Finally, the boundary-condition approach
initiated by Maldacena \cite{ref70} and
further developed in Ref.~\cite{ref74}
addresses aspects of the higher-derivative
ghost problem of Weyl gravity, rather than
establishing the stability of the full
conformal gauge theory prior to symmetry
breaking. A complete Hamiltonian analysis
of the conformal theory, including its
ghost and Ostrogradsky structure and its
continuation to the Weyl phase, lies
outside the scope of the present work
and remains an important direction for
future investigation.

In conclusion, the principal contribution
of the present work is to clarify the
dynamical status of the broken
special-conformal connection in the
gauge-theoretic formulation of conformal
gravity. We have established that this
connection remains auxiliary in the
standard local Yang-Mills-Higgs theory
and demonstrated that, on the source-free
projected branch defined by the vanishing
of its complete additional variation,
its exact elimination dynamically
produces four-dimensional Weyl gravity.
Together with the subsequent scalar
symmetry breaking, these results provide
a systematic framework for investigating
the connection between conformal,
Weyl and Einstein gravities.

\section*{Acknowledgments}
We are grateful to Vasilis Letsios and Pantelis Manousselis for numerous illuminating discussions during the development of the work presented in this paper, as well as Dieter Lust and Anamaria Hell for their positive comments on the manuscript. P.N. acknowledges support from the Rector's Postdoctoral Fellowship Program at Stellenbosch University and the National Institute for Theoretical and Computational Sciences (NITheCS). P.N. also thanks the organizers of the Corfu Summer Institute 2026 for their local support during the workshop, where part of this work was finalized. G.Z. would like to thank the Arnold Sommerfeld Center at LMU Munich and the Max Planck Institute for Physics (MPP) for their hospitality and support, the University of Hamburg and DESY for their hospitality, and the Cluster of Excellence ``Quantum Universe'' and COST Action CA24146 for support. GZ would also like to thank the CERN TH for the hospitality and support, where the present work was finilized.

\appendix
\section{Explicit compact \texorpdfstring{$15$}{15}-plet and \texorpdfstring{$6$}{6}-plet mass spectrum}
\label{app:compact}
The compact real form is useful for displaying the Higgs mass matrix with a positive-definite scalar kinetic normalization. It does not replace the Lorentzian $SU(2,2)$ reality condition used for the physical conformal theory; it provides a controlled representation-theory organization because both real forms share the complexification $\mathfrak{sl}(4,\mathbb C)$.

Let $T_A$ be Hermitian $SU(4)$ generators normalized by
\begin{equation}
\Tr(T_AT_B)=\frac12\delta_{AB}.
\end{equation}
For an adjoint scalar $\Phi=\Phi^AT_A$,
\begin{equation}
D_\mu\Phi=\partial_\mu\Phi-\ii g[A_\mu,\Phi],
\qquad
\mathcal L_{15}=\Tr(D_\mu\Phi D^\mu\Phi).
\end{equation}
Choose
\begin{equation}
\langle\Phi\rangle=v_{15}H,
\qquad
H=\frac{1}{2\sqrt2}\tau_3\otimes\sigma_0,
\qquad
\Tr H^2=\frac12.
\label{eq:adjointVevApp}
\end{equation}
An orthonormal basis adapted to the unbroken subgroup is
\begin{align}
L_i&=\frac14(\tau_0+\tau_3)\otimes\sigma_i,&
R_i&=\frac14(\tau_0-\tau_3)\otimes\sigma_i,\nonumber\\
D&=H,&
V_0&=\frac{1}{2\sqrt2}\tau_2\otimes\sigma_0,\nonumber\\
W_0&=\frac{1}{2\sqrt2}\tau_1\otimes\sigma_0,&
V_i&=\frac{1}{2\sqrt2}\tau_1\otimes\sigma_i,\nonumber\\
W_i&=\frac{1}{2\sqrt2}\tau_2\otimes\sigma_i,&& i=1,2,3.
\label{eq:SU4basisApp}
\end{align}
The six $L_i,R_i$ span $\mathfrak{su}(2)_L\oplus\mathfrak{su}(2)_R$, $D$ is the residual $U(1)$ generator, and the eight $V_a,W_a$ are broken. For comparison with conformal notation define
\begin{equation}
P_a=\frac{V_a+W_a}{\sqrt2},
\qquad
K_a=\frac{V_a-W_a}{\sqrt2}.
\label{eq:PKfromVWApp}
\end{equation}
At the vacuum,
\begin{equation}
D_\mu\langle\Phi\rangle=-\ii gv_{15}A_\mu^A[T_A,H],
\end{equation}
and comparison with $\tfrac12A_\mu^A(M_{15}^2)_{AB}A^{B\mu}$ gives
\begin{equation}
(M_{15}^2)_{AB}=2g^2v_{15}^2\Tr([T_A,H]^\dagger[T_B,H]).
\label{eq:M15App}
\end{equation}
The generators $L_i,R_i,D$ commute with $H$ and remain massless, while every $V_a,W_a$ has
\begin{equation}
\Tr([T,H]^\dagger[T,H])=\frac14,
\end{equation}
so
\begin{equation}
m_{V_a}^2=m_{W_a}^2\equiv a=\frac{g^2v_{15}^2}{2}.
\label{eq:stageAmassApp}
\end{equation}
Hence the first vacuum produces seven massless and eight degenerate massive gauge directions. This degeneracy means that $V/W$ and $P/K$ are equivalent bases of the same broken eigenspace at this stage; the second vacuum is what selects the final combinations.

For the antisymmetric $\mathbf6$, represent the scalar by $\Sigma=-\Sigma^T$ transforming as $\Sigma\to U\Sigma U^T$, with
\begin{equation}
D_\mu\Sigma=\partial_\mu\Sigma-\ii g(A_\mu\Sigma+\Sigma A_\mu^T),
\qquad
\mathcal L_6=\frac12\Tr[(D_\mu\Sigma)^\dagger D^\mu\Sigma].
\end{equation}
Choose
\begin{equation}
\langle\Sigma\rangle=\frac{v_6}{2}\Omega,
\qquad
\Omega=\tau_0\otimes\varepsilon,
\qquad
\varepsilon=\ii\sigma_2.
\label{eq:6vevApp}
\end{equation}
With $Q_A=T_A\Omega+\Omega T_A^T$,
\begin{equation}
(M_6^2)_{AB}=\frac{g^2v_6^2}{4}\Tr(Q_A^\dagger Q_B).
\end{equation}
Using $\sigma_i\varepsilon+\varepsilon\sigma_i^T=0$ gives
\begin{align}
M_6^2(L_i)&=M_6^2(R_i)=M_6^2(V_a)=0,\nonumber\\
M_6^2(D)&=\frac{g^2v_6^2}{2},
\qquad
M_6^2(W_a)=\frac{g^2v_6^2}{2}.
\end{align}
With
\begin{equation}
a=\frac{g^2v_{15}^2}{2},
\qquad
c=\frac{g^2v_6^2}{2},
\end{equation}
the complete eigenvalue pattern is
\begin{equation}
0\ (6),\qquad a\ (4),\qquad c\ (1),\qquad a+c\ (4).
\label{eq:fullSpectrumApp}
\end{equation}
In the $(P_a,K_a)$ basis the identical $2\times2$ blocks are
\begin{equation}
M_{PK}^2=\begin{pmatrix}
a+c/2&-c/2\\
-c/2&a+c/2
\end{pmatrix},
\end{equation}
with eigenvectors
\begin{equation}
\frac{P_a+K_a}{\sqrt2}=V_a,\qquad m_V^2=a,
\qquad
\frac{P_a-K_a}{\sqrt2}=W_a,\qquad m_W^2=a+c.
\end{equation}
Thus the $15$-plet vacuum identifies the eight broken conformal directions, while the $6$-plet vacuum resolves the final mass eigenbasis and breaks the residual $U(1)$ direction. This compact calculation should be interpreted as symmetry-breaking bookkeeping; the Lorentzian propagating content is fixed separately by the invariant bilinear and the physical reality conditions.

\section{Real-form and complex-basis dictionary}
The physical Lorentzian gauge algebra is $\mathfrak{so}(2,4)\simeq\mathfrak{su}(2,2)$, whereas the compact Higgs calculation uses $\mathfrak{su}(4)$. They are inequivalent real algebras but share the same complexification,
\begin{equation}
\mathfrak{su}(4)_\mathbb C\simeq\mathfrak{su}(2,2)_\mathbb C\simeq\mathfrak{sl}(4,\mathbb C).
\end{equation}
The compact condition is $X^\dagger+X=0$, whereas the noncompact reality condition is
\begin{equation}
X^\dagger\eta_{2,2}+\eta_{2,2}X=0,
\qquad \eta_{2,2}=\mathrm{diag}(1,1,-1,-1).
\end{equation}
Analytic continuation therefore preserves the complex commutator algebra and the representation decomposition, but not the positive-definite quadratic form of the compact theory. This is why compact Higgs masses and Lorentzian kinetic signs must not be identified without the appropriate reality condition.

Splitting the complex fundamental gives
\begin{equation}
4=(2,1)_{+1}\oplus(1,2)_{-1},
\end{equation}
so
\begin{equation}
15=(3,1)_0\oplus(1,3)_0\oplus(1,1)_0\oplus(2,2)_{+2}\oplus(2,2)_{-2}.
\end{equation}
The two complex $(2,2)$ blocks are the representation-theory home of the coordinates $Q_a,Q_a^\star$ introduced in the main text. The corresponding one-forms $Z^a,Z^{\star a}$ are merely complex coordinates on the real physical pair $(e^a,b^a)$; they do not double the gauge-field content.

\section{Component derivation of the projected epsilon contractions}
\label{app:epsilon}
For completeness we record the algebra behind the exact Stage-A reduction. With
\begin{equation}
X_{\mu\nu}{}^{ab}=\frac14\left(e_\mu{}^ab_\nu{}^b-e_\mu{}^bb_\nu{}^a-e_\nu{}^ab_\mu{}^b+e_\nu{}^bb_\mu{}^a\right),
\end{equation}
and $\epsilon^{0123}=+1$, $\epsilon_{0123}=-1$, the pure curvature contraction gives
\begin{equation}
\frac14\epsilon^{\mu\nu\rho\sigma}\epsilon_{abcd}R_{\mu\nu}{}^{ab}R_{\rho\sigma}{}^{cd}
=-\sqrt{-g}\,\mathcal E_4.
\end{equation}
For the mixed contraction, antisymmetry makes the four terms in $X_{\rho\sigma}{}^{cd}$ equal after relabeling. The vierbein form of the spacetime epsilon tensor and the generalized Kronecker delta then yield
\begin{equation}
\epsilon^{\mu\nu\rho\sigma}\epsilon_{abcd}R_{\mu\nu}{}^{ab}e_\rho{}^ce^{\lambda d}
=4\sqrt{-g}\left(R^{\sigma\lambda}-\frac12Rg^{\sigma\lambda}\right).
\label{eq:epsilonMixedApp}
\end{equation}
Multiplication by $b_{\sigma\lambda}$ gives Eq.~\eqref{eq:epsRX}.

The quadratic contraction is
\begin{equation}
\epsilon^{\mu\nu\rho\sigma}\epsilon_{abcd}X_{\mu\nu}{}^{ab}X_{\rho\sigma}{}^{cd}
=2\sqrt{-g}\left(b_{\mu\nu}b^{\nu\mu}-b^2\right).
\end{equation}
Writing $b_{\mu\nu}=s_{\mu\nu}+a_{\mu\nu}$ with $s_{\mu\nu}=s_{\nu\mu}$ and $a_{\mu\nu}=-a_{\nu\mu}$ gives
\begin{equation}
b_{\mu\nu}b^{\nu\mu}-b^2
=s_{\mu\nu}s^{\mu\nu}-s^2-a_{\mu\nu}a^{\mu\nu}.
\end{equation}
The mixed term couples only to $s_{\mu\nu}$ because $G_{\mu\nu}$ is symmetric, whereas variation of the final term sets $a_{\mu\nu}=0$. Thus the symmetric Fierz--Pauli trace combination emerges dynamically from the unrestricted tensor equation rather than from an assumption of symmetry.

\section{Linearized identities on the de Sitter background}
\label{app:dSidentities}
For reference we collect the linearized curvature identities used in Sec.~\ref{sec:dS}. Let
\begin{equation}
g_{\mu\nu}=\bar g_{\mu\nu}+h_{\mu\nu},
\qquad h=\bar g^{\mu\nu}h_{\mu\nu}.
\end{equation}
The variation of the Christoffel connection is
\begin{equation}
\delta\Gamma^\rho{}_{\mu\nu}
=\frac12\bar g^{\rho\sigma}
(\bar\nabla_\mu h_{\nu\sigma}+\bar\nabla_\nu h_{\mu\sigma}-\bar\nabla_\sigma h_{\mu\nu}),
\end{equation}
and therefore
\begin{align}
R^L_{\mu\nu}
=\frac12(&\bar\nabla_\rho\bar\nabla_\mu h^\rho{}_{\nu}
+\bar\nabla_\rho\bar\nabla_\nu h^\rho{}_{\mu}
-\bar\square h_{\mu\nu}
-\bar\nabla_\mu\bar\nabla_\nu h).
\label{eq:RicciLinearApp}
\end{align}
The scalar curvature variation is
\begin{equation}
R^L=\bar\nabla_\mu\bar\nabla_\nu h^{\mu\nu}-\bar\square h-\Lambda h.
\end{equation}
The linearized cosmological Einstein tensor is
\begin{equation}
\mathcal G^L_{\mu\nu}=R^L_{\mu\nu}-\frac12\bar g_{\mu\nu}R^L-\Lambda h_{\mu\nu}.
\label{eq:GLgeneralApp}
\end{equation}
It obeys the linearized Bianchi identity
\begin{equation}
\bar\nabla^\mu\mathcal G^L_{\mu\nu}=0.
\end{equation}

For a transverse-traceless fluctuation, $\bar\nabla^\mu h_{\mu\nu}=0$ and $h=0$. Commuting covariant derivatives on the de Sitter background gives
\begin{equation}
R^L_{\mu\nu}[h^{\rm TT}]
=-\frac12\bar\square h^{\rm TT}_{\mu\nu}
+\frac{4\Lambda}{3}h^{\rm TT}_{\mu\nu},
\end{equation}
while $R^L=0$. Consequently,
\begin{align}
\mathcal G^L_{\mu\nu}[h^{\rm TT}]
&=R^L_{\mu\nu}-\Lambda h^{\rm TT}_{\mu\nu}\\
&=-\frac12\left(\bar\square-\frac{2\Lambda}{3}\right)h^{\rm TT}_{\mu\nu},
\end{align}
which is Eq.~\eqref{eq:GTTfull}.

The auxiliary fluctuation equation
\begin{equation}
\mathcal G^L_{\mu\nu}-4(\phi_{\mu\nu}-\bar g_{\mu\nu}\phi)=0
\end{equation}
has trace
\begin{equation}
\mathcal G^L+12\phi=0,
\end{equation}
so
\begin{equation}
\phi=-\frac{1}{12}\mathcal G^L.
\end{equation}
Substitution then yields
\begin{equation}
\phi_{\mu\nu}=\frac14\left(\mathcal G^L_{\mu\nu}-\frac13\bar g_{\mu\nu}\mathcal G^L\right),
\end{equation}
which directly produces the curvature-squared quadratic action after elimination.

In TT gauge, Eq.~\eqref{eq:quadraticEffFull} becomes
\begin{align}
S_{\rm eff}^{(2)}
=-2a_{\CG}\int\dd^4x\sqrt{-\bar g}
\left[\mathcal G^L_{\mu\nu}\mathcal G_L^{\mu\nu}
+\frac{\Lambda}{3}h_{\mu\nu}\mathcal G_L^{\mu\nu}\right].
\end{align}
Using
\begin{equation}
\mathcal G^L_{\mu\nu}=-\frac12L_0h_{\mu\nu},
\qquad L_0=\bar\square-\frac{2\Lambda}{3},
\end{equation}
and integrating by parts gives
\begin{align}
S_{\rm TT}^{(2)}
&=-\frac{a_{\CG}}{2}\int\dd^4x\sqrt{-\bar g}\,
h^{\mu\nu}L_0\left(L_0-\frac{2\Lambda}{3}\right)h_{\mu\nu}\\
&=-\frac{a_{\CG}}{2}\int\dd^4x\sqrt{-\bar g}\,
h^{\mu\nu}
\left(\bar\square-\frac{2\Lambda}{3}\right)
\left(\bar\square-\frac{4\Lambda}{3}\right)h_{\mu\nu}.
\end{align}
This yields the factorized equation in the main text.

The two projector formulas can also be verified directly. Let
\begin{equation}
h^{(0)}=-\frac{3}{2\Lambda}L_{\PM}h,
\qquad
h^{(\PM)}=\frac{3}{2\Lambda}L_0h,
\end{equation}
with $L_{\PM}=L_0-2\Lambda/3$. Then
\begin{align}
h^{(0)}+h^{(\PM)}
&=\frac{3}{2\Lambda}(L_0-L_{\PM})h\\
&=h.
\end{align}
Moreover,
\begin{equation}
L_0h^{(0)}=-\frac{3}{2\Lambda}L_0L_{\PM}h=0,
\end{equation}
and similarly $L_{\PM}h^{(\PM)}=0$. The factorization therefore defines an explicit decomposition of any TT solution into the two spin--2 branches whenever $\Lambda\neq0$.

\section{Scope of the auxiliary theorem}
The result proved in the main text concerns the fundamental local action with $(e,\omega,b,\chi,\Phi_r)$ retained. It is useful to state precisely when the conclusion can fail.

First, an explicit operator
\begin{equation}
\Delta S_K=\alpha_K\int\dd^4x\sqrt{-g}\,S_{\mu\nu}{}^aS^{\mu\nu}{}_a,
\qquad \alpha_K\neq0,
\end{equation}
contains
\begin{equation}
S_{\mu\nu}{}^a=2\partial_{[\mu}b_{\nu]}{}^a+\cdots
\end{equation}
and hence a genuine $(\partial b)^2$ term. The $b$ equation then contains second derivatives and the field is no longer auxiliary.

Second, a generalized effective theory may contain a scalar-dependent gauge-kinetic tensor,
\begin{equation}
S_G=-\frac14\int\dd^4x\sqrt{-g}\,G_{AB}(\Phi)F_{\mu\nu}^AF^{B\mu\nu}.
\end{equation}
If $G_{AB}(\Phi)$ is not the constant adjoint-invariant bilinear, symmetry breaking may generate a nonzero effective $K$--$K$ block and therefore an $S^2$ operator. This is an additional effective-field-theory assumption rather than a consequence of the ordinary Yang--Mills contraction considered here.

Third, if another dynamical field is integrated out before $b_\mu{}^a$, the resulting effective action can be nonlocal. Such a nonlocal action may exhibit derivative dependence on $b$ even though the original local theory had an algebraic $b$ equation after integration by parts. This does not contradict the local auxiliary theorem; it reflects the changed order of field elimination.

These qualifications delimit the claim. The theorem is not that every possible conformal effective action makes $b_\mu{}^a$ auxiliary. It is that the ordinary local $SO(2,4)$ Yang--Mills--Higgs contraction, together with the projected gravitational term and standard scalar kinetic sectors used here, does not generate an independent $F^{(K)2}$ kinetic block.

\section{Bispinor form of the reduced auxiliary equation}
\label{app:bispinor}
The complex Lorentz representation also provides a useful check of the source-free projected tensor reduction. Introduce the soldering matrices $\sigma^a_{\alpha\dot\alpha}$ and write
\begin{equation}
V_{\alpha\dot\alpha}=\sigma^a_{\alpha\dot\alpha}V_a,
\qquad
f_{\mu\,\alpha\dot\alpha}=\sigma^a_{\alpha\dot\alpha}b_{\mu a}.
\end{equation}
Using the vierbein to convert the spacetime index,
\begin{equation}
b_{\alpha\dot\alpha,\beta\dot\beta}
\equiv e_{\beta\dot\beta}{}^\mu f_{\mu\,\alpha\dot\alpha}.
\label{eq:bispinorB}
\end{equation}
No symmetry under exchange of the two spinor pairs is assumed before variation. In the normalization corresponding to the tensor equation used in the main text, the reduced $b$-dependent action can be displayed as
\begin{align}
S_A={}&-16a_{\CG}\int\dd^4x\sqrt{-g}\Big[
G^{\alpha\dot\alpha,\beta\dot\beta}
 b_{\alpha\dot\alpha,\beta\dot\beta}
-2b_{\alpha\dot\alpha,\beta\dot\beta}
 b^{\beta\dot\beta,\alpha\dot\alpha}
+2b^2\Big],
\label{eq:bispinorAction}
\end{align}
where
\begin{equation}
b\equiv g^{\alpha\dot\alpha,\beta\dot\beta}
 b_{\alpha\dot\alpha,\beta\dot\beta}.
\end{equation}
Variation gives
\begin{equation}
G_{\alpha\dot\alpha,\beta\dot\beta}
-4b_{\beta\dot\beta,\alpha\dot\alpha}
+4g_{\alpha\dot\alpha,\beta\dot\beta}b=0.
\label{eq:bispinorEom}
\end{equation}
Because both the Einstein tensor and the metric are symmetric under exchange of the two vector-index pairs, the antisymmetric part of Eq.~\eqref{eq:bispinorEom} yields
\begin{equation}
b_{[\alpha\dot\alpha,\beta\dot\beta]}=0.
\end{equation}
Taking the trace in four dimensions gives
\begin{equation}
G+12b=0.
\end{equation}
Since $G=-R$,
\begin{equation}
b=\frac{R}{12}.
\end{equation}
Substituting back,
\begin{align}
b_{\alpha\dot\alpha,\beta\dot\beta}
&=\frac14\left(
G_{\alpha\dot\alpha,\beta\dot\beta}
-\frac13g_{\alpha\dot\alpha,\beta\dot\beta}G
\right)\\
&=\frac14\left(
R_{\alpha\dot\alpha,\beta\dot\beta}
-\frac16R g_{\alpha\dot\alpha,\beta\dot\beta}
\right)\\
&=\frac12P_{\alpha\dot\alpha,\beta\dot\beta}.
\label{eq:bispinorSchouten}
\end{align}
This is the spinorial form of Eq.~\eqref{eq:bSchouten} on the source-free projected branch. The result is useful because the $(2,2)$ representation of $SU(2)_L\times SU(2)_R$ is manifest: the special-conformal connection is converted into a bispinor carrying exactly the vector representation, while its pair-exchange antisymmetric component is removed by the projected algebraic field equation.

The physical significance is unchanged by the notation. Complex spinor indices reorganize the Lorentz representation but do not create additional complex degrees of freedom. On the source-free projected branch, the $SU(2,2)$ reality condition maps the spinor expression back to the real Lorentzian tensor $b_{\mu\nu}$ fixed by the real Schouten tensor.

\section{Notation, conventions, and dimensional checks}
\label{app:notation}
For clarity we collect the conventions used throughout the derivations. Spacetime indices are $\mu,\nu,\rho,\sigma=0,1,2,3$, Lorentz indices are $a,b,c,d=0,1,2,3$, and spinor indices are $\alpha,\beta=1,2$ and $\dot\alpha,\dot\beta=1,2$. The metric signature is mostly plus,
\begin{equation}
\eta_{ab}=\mathrm{diag}(-,+,+,+),
\end{equation}
with
\begin{equation}
g_{\mu\nu}=e_\mu{}^ae_{\nu a},
\qquad e\equiv\det(e_\mu{}^a)=\sqrt{-g}.
\end{equation}
The orientation is fixed by
\begin{equation}
\epsilon^{0123}=+1,
\qquad
\epsilon_{0123}=-1.
\end{equation}
Normalized antisymmetrization is used,
\begin{equation}
X_{[\mu\nu]}\equiv\frac12(X_{\mu\nu}-X_{\nu\mu}),
\qquad
Y^{[ab]}\equiv\frac12(Y^{ab}-Y^{ba}).
\end{equation}
Thus
\begin{equation}
e_{[\mu}{}^{[a}b_{\nu]}{}^{b]}
=\frac14\left(e_\mu{}^ab_\nu{}^b-e_\mu{}^bb_\nu{}^a-e_\nu{}^ab_\mu{}^b+e_\nu{}^bb_\mu{}^a\right).
\end{equation}

The Lorentz-covariant exterior derivative acts as
\begin{equation}
D^{(\omega)}e^a=\dd e^a+\omega^a{}_b\wedge e^b,
\qquad
D^{(\omega)}b^a=\dd b^a+\omega^a{}_b\wedge b^b.
\end{equation}
The physical curvatures are
\begin{align}
T^a&=D^{(\omega)}e^a+\chi\wedge e^a,\\
S^a&=D^{(\omega)}b^a-\chi\wedge b^a,\\
F_D&=\dd\chi+2e^a\wedge b_a,\\
R^{ab}&=R^{ab}(\omega)-4e^{[a}\wedge b^{b]}.
\end{align}
Only $S^a$ contains $\dd b^a$.

The spacetime tensor associated with the special-conformal connection is
\begin{equation}
b_{\mu\nu}=e_{\nu a}b_\mu{}^a,
\qquad
b=g^{\mu\nu}b_{\mu\nu}.
\end{equation}
No symmetry of $b_{\mu\nu}$ is assumed before variation. On the source-free projected branch its on-shell symmetry follows from the antisymmetric part of the projected field equation.

The four-dimensional Schouten and Weyl tensors are
\begin{equation}
P_{\mu\nu}=\frac12\left(R_{\mu\nu}-\frac16Rg_{\mu\nu}\right),
\end{equation}
and
\begin{align}
C_{\mu\nu\rho\sigma}
={}&R_{\mu\nu\rho\sigma}
-2\left(g_{\mu[\rho}P_{\sigma]\nu}-g_{\nu[\rho}P_{\sigma]\mu}\right).
\end{align}
The curvature-squared identities used repeatedly are
\begin{equation}
C_{\mu\nu\rho\sigma}C^{\mu\nu\rho\sigma}
=R_{\mu\nu\rho\sigma}R^{\mu\nu\rho\sigma}
-2R_{\mu\nu}R^{\mu\nu}+\frac13R^2,
\end{equation}
and
\begin{equation}
\mathcal E_4
=R_{\mu\nu\rho\sigma}R^{\mu\nu\rho\sigma}
-4R_{\mu\nu}R^{\mu\nu}+R^2.
\end{equation}
Consequently,
\begin{equation}
C^2=\mathcal E_4+2\left(R_{\mu\nu}R^{\mu\nu}-\frac13R^2\right),
\end{equation}
which is the identity used in the exact Stage-A reduction.

In the complex $P$--$K$ basis, the symbol $\star$ is a reality-conjugation label,
\begin{equation}
Q_a^\star=\frac{P_a+\ii K_a}{\sqrt2},
\qquad
Z^{\star a}=\frac{e^a-\ii b^a}{\sqrt2},
\end{equation}
and must not be confused with the Hodge dual acting on differential forms. The Hodge dual appears only through expressions such as $F\wedge\star F$ in the Yang--Mills action.

The normalizations of $e^a$ and $b^a$ are convention dependent because rescalings of the translation and special-conformal generators can be compensated by inverse rescalings of their gauge fields. All equations in this paper therefore use the fixed generator normalization defined by the commutators in Eq.~\eqref{eq:alg}. The invariant dimensional statement needed for the de Sitter spectrum is
\begin{equation}
[\Lambda]=[m_{\PM}^2],
\qquad
m_{\PM}^2=\frac{2\Lambda}{3}.
\end{equation}
The Higgs scales $gv_{15}$ and $gv_6$ and the geometric curvature scale $\sqrt\Lambda$ remain conceptually distinct until a scalar potential and a complete vacuum solution relate them dynamically.

\section{Metric and its relation to the gauge fields}
\label{app:metric}
\paragraph{Before the second breaking.}
In the action~\eqref{eq:fullActionIntro}, only $S_{\YM}$ and $S_\Phi$ require a metric, while $S_{\rm MM}$ does not. The adjoint vacuum $\langle\Phi_{15}\rangle\propto D$ leaves the subgroup $SO(1,3)\times D$ unbroken and splits the connection into the broken components $e^a$ and $b^a$, which are the weight $+1$ and weight $-1$ eigenspaces of $\mathrm{ad}_D$, since $[D,P_a]=P_a$ and $[D,K_a]=-K_a$. The Lorentz-invariant quadratic forms in these components have dilatation weights
\begin{equation}
e\!\cdot\! e:\ +2,\qquad e\!\cdot\! b:\ 0,\qquad b\!\cdot\! b:\ -2 .
\end{equation}
In the Weyl phase $D$ is unbroken, so a metric must carry weight $+2$ for the Yang--Mills and Weyl terms to be consistent with local scale transformations. The only such form is $g_{\mu\nu}=\eta_{ab}e_\mu{}^ae_\nu{}^b$. It is not invariant under the full gauge group, since $\kappa(P_a,P_b)=0$ prevents any $\SO(2,4)$-invariant bilinear from defining a metric on the $e$-directions; the Lorentz-invariant tensor $\eta_{ab}$ is used on them, and the form is invariant under special-conformal transformations ($\delta_Ke^a=0$) but not under local translations.

The invariant bilinear does produce one symmetric tensor from both fields,
\begin{equation}
\kappa(A_\mu^{\rm broken},A_\nu^{\rm broken})=2c\,e_{(\mu}{}^{a}b_{\nu)a}=2c\,b_{(\mu\nu)} ,
\end{equation}
with $c$ from Eq.~\eqref{eq:PKpair}. It has weight $0$ and is therefore not a metric. On the source-free projected branch it equals $\tfrac12P_{\mu\nu}[g]$, and on an Einstein-type vacuum $b_\mu{}^a=c\,e_\mu{}^a$ it is proportional to $g_{\mu\nu}$.

\paragraph{After the second breaking.}
With both vacua, the mass eigenstates are $V_a=(P_a+K_a)/\sqrt2$ and $W_a=(P_a-K_a)/\sqrt2$ with $m_V^2=a$ and $m_W^2=a+c$ (Appendix~\ref{app:compact}). Since
\begin{equation}
e^aP_a+b^aK_a=\frac{e^a+b^a}{\sqrt2}\,V_a+\frac{e^a-b^a}{\sqrt2}\,W_a ,
\end{equation}
the light frame field is
\begin{equation}
\tilde e_\mu{}^a=\frac{e_\mu{}^a+b_\mu{}^a}{\sqrt2},
\qquad
\tilde g_{\mu\nu}=\eta_{ab}\,\tilde e_\mu{}^a\tilde e_\nu{}^b ,
\end{equation}
while $(e^a-b^a)/\sqrt2$ is the heavy field. Mixing $e$ and $b$ is allowed here because the $6$-plet vacuum gives $D$ a mass ($M_6^2(D)=c$), so dilatation weight is no longer a symmetry. At energies well below the heavy mass, $E^2\ll m_W^2=a+c$, the heavy field $(e_\mu{}^a-b_\mu{}^a)/\sqrt2$ can be integrated out and set to zero, which gives the proportional form $b_\mu{}^a\propto e_\mu{}^a$ of Sec.~2.3.

\end{document}